\documentclass{emulateapj}
\usepackage{apjfonts}

\providecommand{\unit}[1]{\ensuremath{\:\mathrm{#1}}}

\def\eg{{e.g.}\ }

\def\ie{{i.e.}\ }
\def\vs{{vs.}\ }

\newcommand{\Bo}{\ensuremath{\vec{B}_{\parallel}}}
\newcommand{\dz}{\ensuremath{\delta z}}
\newcommand{\Exp}[1]{\ensuremath{\mathrm{e}^{#1}}}
\newcommand{\tamin}{\ensuremath{\tau_{A,\mathrm{min}}}}
\newcommand{\tamax}{\ensuremath{\tau_{A,\mathrm{max}}}}
\newcommand{\tnl}{\ensuremath{\tau_{\mathrm{NL}}}}
\newcommand{\tnlmin}{\ensuremath{\tau_{\mathrm{NL},\mathrm{min}}}}

\newcommand{\tnu}{\ensuremath{\tau_{\nu}}}
\newcommand{\tnumin}{\ensuremath{\tau_{\nu,\mathrm{min}}}}

\newcommand{\shellatm}{\textsc{Shell-Atm}}

\shorttitle{Shell-model of RMHD turbulence and coronal heating}
\shortauthors{Buchlin and Velli}

\begin{document}

\title{Shell-models of RMHD turbulence and the heating of solar coronal
  loops}

\author{E. Buchlin\altaffilmark{1,2}}
\author{M. Velli\altaffilmark{3,1}}

\altaffiltext{1}{ Dipartimento di Astronomia e Scienza dello Spazio,
  Universit\`a di Firenze, Largo E. Fermi 2, 50125 Firenze, Italy }

\altaffiltext{2}{Space and Atmospheric Physics Department, The Blackett
  Laboratory, Imperial College, London SW7 2BW, UK}

\altaffiltext{3}{ Jet Propulsion Laboratory, California Institute of
  Technology, 4800 Oak Grove Drive, Pasadena, CA 91109, USA }

\begin{abstract}
  A simplified non-linear numerical model for the development of
  incompressible magnetohydrodynamics (MHD) in the presence of a strong
  magnetic field \Bo\ and stratification, nicknamed \shellatm, is presented.
  In planes orthogonal to the mean field, the non-linear incompressible
  dynamics is replaced by 2D shell-models for the complex variables $u$ and
  $b$, allowing one to reach large Reynolds numbers while at the same time
  carrying out sufficiently long time integrations to obtain a good
  statistics at moderate computational cost.  The shell-models of different
  planes are coupled by Alfv\'en waves propagating along \Bo. The model may
  be applied to open or closed magnetic field configurations where the axial
  field dominates and the plasma pressure is low; here we apply it to the
  specific case of a magnetic loop of the solar corona heated via turbulence
  driven by photospheric motions, and we use statistics for its analysis.
  The Alfv\'en waves interact non-linearly and form turbulent spectra in the
  directions perpendicular and, via propagation, also parallel to the mean
  field.  A heating function is obtained, and is shown to be intermittent;
  the average heating is consistent with values required for sustaining a
  hot corona, and is proportional to the aspect ratio of the loop to the
  power $-1.5$; characteristic properties of heating events are distributed
  as power-laws. Cross-correlations show a delay of dissipation compared to
  energy content.
\end{abstract}

\keywords{
  Sun\,: corona, flares -- MHD -- turbulence
}

\section{Introduction}

MHD turbulence in the presence of a mean magnetic field, with or without
density and gravitational gradients, plays a role in many environments,
ranging from stellar coronae and winds \citep{kleinlw91} to the interstellar
medium \citep{desai94} and accretion disks. In such regions, energy may be
transferred, accumulated and dissipated in a way which is inherently
anisotropic \citep{sheba83,oug94,kin98,muller03,oug04}.

In particular, in solar coronal physics, where one of the main problems is
to understand how the corona can be sustained at more than a million of
degrees, it is believed that the necessary heating could be produced at
small scales generated by a non-linear cascade along a turbulent spectrum
\citep{hey92,gom92}. Furthermore, as flux tubes (\eg in the form of coronal
loops or coronal funnels) are omnipresent, the anisotropy coming from the
dominant magnetic field may be a central feature of the processes governing
energy dissipation, like the non-linear collisions of counter-propagating
Alfv\'en wave packets. It can thus be expected that solving the coronal
heating problem, \ie understanding how the temperature of the corona can be
sustained, may require to understand the details of the turbulent dynamics
of MHD in these environments.

One way to study and the dynamics of such system is to perform direct
numerical simulations (DNS).  In the case of anisotropic MHD, DNS have
provided insight on subjects like the anisotropy of the spectra
\citep[\eg][]{kin98,oug04}, the parametric decay of Alfv\'en waves
\citep[\eg][]{delz01}, and Alfv\'en waves filamentation
\citep[\eg][]{passot03}.  MHD simulations are also used to study the
topology of magnetic field lines and magnetic reconnection in the corona
\citep[\eg][]{aulanier05b}. But the Reynolds numbers attained in all the DNS
up to now are below $10^3$, while they are believed to be $10^{12}$ to
$10^{14}$ in the corona.  DNS are very far from being able to represent all
the scales of turbulence in the corona, there is a huge gap to fill.
Furthermore, as statistics are of great help in the study of turbulence,
attempts have been made to analyze statistically energy dissipations
produced by DNS.  Distributions of events are for instance presented in
\cite{dmi98} and \cite{gev98} from 2D DNS of reduced MHD.  But it is still
difficult to get significant statistics from 3D DNS, and it is even more
difficult when trying to go to higher Reynolds numbers because then the grid
resolution must be higher and the computations of the model are too slow.
For all these reasons, there is a need for simplified numerical models of
MHD, which would run sufficiently fast to get statistics of turbulence at
high Reynolds numbers, while keeping the most relevant features of MHD
turbulence.

Several approaches have been used to build such simplified numerical models
of MHD. For example, the Self-Organized Criticality behavior (SOC) of MHD
systems can be modeled by cellular automata (CA), where the interactions of
individual cells translate into a global statistic behavior of the whole
system, following the first models of \citet{luh91,luh93}. However, the need
for physical realism is not entirely addressed by the cellular automata,
despite efforts trying to include the constraints issued from the MHD
equations \citep{vla95, isl00, isl01, buc03}.

Another approach is to simplify the non-linear interactions by reducing the
number of modes which are allowed to interact non-linearly. In the coronal
loop context, a shell-model approach has been used by \citet{nig04,nig05}.
We have developed a similar numerical model independently, starting from the
reduced MHD equations but allowing for the stratification of the plasma.
This numerical code, nicknamed \shellatm, allows one to reach (kinetic and
magnetic) Reynolds numbers unachieved before. In this paper, we will focus
on the problem of a coronal loop where energy is forced into the system by
footpoint motions, describing in detail the dynamics of the heating events,
turbulence spectra, statistics and scaling laws.

\section{Description of the \shellatm\ model}
\label{sec:desc}

We start from an approximation to incompressible MHD known as reduced MHD
\citep[RMHD: ][]{kadomtsev74,str76}, which is valid when the plasma $\beta$
parameter (kinetic over magnetic pressure) is low, the domain has a large
aspect ratio ($a = \ell / L \ll 1$) and the poloidal field is small compared
to a strong axial external \Bo\ magnetic field ($B_{\perp} / B_{\parallel}
\la a$). In this approximation the largest extension $L$ of the domain
defines the parallel direction, or $z$-axis; the velocity field is only
composed of fluctuations $\vec u_{\perp}$ orthogonal to the $z$-axis; the
magnetic field can be decomposed into $\vec B_{\parallel} + \vec B_{\perp}$,
where $\vec B_{\parallel} = B_{\parallel} \hat e_z$ is the average magnetic
field, parallel to the $z$-axis, and $\vec B_{\perp}$ is a perpendicular
fluctuation.  Throughout, we will normalize the magnetic fields by
$\sqrt{\mu_0\rho_0}$, considering for the moment a medium with uniform
density $\rho_0$ ($\vec b_{\parallel} = \vec B_{\parallel} /
\sqrt{\mu_0\rho_0}$ and $\vec b_{\perp} = \vec B_{\perp} /
\sqrt{\mu_0\rho_0}$).  The equations of RMHD become:
\begin{eqnarray}
  \label{eq:rmhd1}
  \frac{\partial \vec u_{\perp}}{\partial t}
  + \vec u_{\perp} \cdot \vec \nabla \vec u_{\perp}
  & = & - \vec \nabla _{\perp} \left(\frac{p}{\rho_0} +
    \frac12 b^2_{\perp}\right) 
  + \vec b_{\perp} \cdot \vec \nabla \vec b_{\perp} \nonumber \\
  && \qquad\qquad + b_{\parallel} \frac{\partial \vec b_{\perp}}{\partial z} +
  \nu \nabla^2_{\perp} \vec u_{\perp}\\ 
  \label{eq:rmhd2}
  \frac{\partial \vec b_{\perp}}{\partial t} &= &
  \vec b_{\perp} \cdot \vec \nabla \vec u_{\perp} -
  \vec u_{\perp} \cdot \vec \nabla \vec b_{\perp} \nonumber \\
  && \qquad\qquad + b_{\parallel} \frac{\partial \vec u_{\perp}}{\partial z}
  + \eta \nabla^2_{\perp} \vec b_{\perp}  \\
  \label{eq:rmhd3}
  \nabla \cdot \vec u_{\perp} = 0 &&\qquad\quad \nabla \cdot \vec b_{\perp}
  = 0
\end{eqnarray}

As one can see in these equations, the non-linear dynamics occurs only in
the planes perpendicular to the mean field \Bo\ while Alfv\'en waves
propagate along the mean field. Direct simulations of these equations in one
plane \citep{dmi98,gev98} or in a 3D box \citep{dmi03b} have been carried
out but the Reynolds numbers obtained with such simulations are much too low
to get a realistic inertial range of turbulence and long-term statistics. It
is therefore our interest to simplify this model further by reducing the
dynamics in the planes. This can be done by using shell-models, as described
below.

The plasma of the solar corona and solar wind is stratified, so that one
must allow for gradients of the mass density $\rho$ even while considering
incompressible couplings.  Stratification couples incompressible Alfv\'en
waves by introducing variations in Alfv\'en speed and therefore reflection
(as well as amplification/depression of amplitudes due to the conservation
of energy flux).  Such terms may be written more clearly in terms of the the
Els\"asser variables $\vec Z^\pm = \vec u_{\perp} \pm \vec b_{\perp}$ (with
$\vec b_{\perp} = \vec B_{\perp} / \sqrt{\mu_0\rho}$), in which case the
effect of stratification on the linear propagation of modes may be written
as \citep{velli93}:
\begin{equation}
  \label{eq:velli93}
  \frac{\partial\vec Z^\pm}{\partial t} \pm \vec b_{\parallel}
  \cdot \vec \nabla \vec Z^\pm  \mp \vec Z^\mp \cdot \vec \nabla
  \vec b_{\parallel} \pm \frac12 \left( \vec Z^\mp - \vec Z^\pm \right)
  \vec \nabla \cdot \vec b_{\parallel} = 0
\end{equation}

The first two terms describe the wave propagation, the third term the
reflection of the waves by the perpendicular gradient of the Alfv\'en speed
(which vanishes for a non-diverging flux tube), the fourth term describes
the growth or decrease in the normalized wave amplitude due to variations in
Alfv\'en speed -- assuring conservation of wave energy flux -- as well as
the isotropic part of the reflection.  We will incorporate these terms in
the general framework of Eqs.~(\ref{eq:rmhd1}-\ref{eq:rmhd3}), but first we
discuss how the non-linear couplings are modeled in the shell approximation.

\subsection{Classical MHD shell-models}

In shell-models of incompressible MHD turbulence \citep{glo85, bis94, giu98,
  bof99, giu02}, one starts by taking the Fourier transform of the MHD
equations and dividing wavevector space into concentric shells $S_n =
\{\vec{k} \;|\; \|\vec{k}\| \in [k_n, k_{n+1}]\}$ with $k_n = k_0 \lambda ^
n $, $n=0,\ldots,n_{\perp}-1$, and usually $\lambda=2$. Also, a single
complex scalar value $u_n$ is chosen to represent the original longitudinal
velocity increments $\left(\vec u(\vec x+\vec\ell)- \vec u(\vec
  x)\right)\cdot \vec\ell / \ell$ on scales $\ell$ for $2\pi/\ell \in S_n$.
The same approximation is made for the magnetic field: a scalar value $b_n$
represents the magnetic field increments on the same scales $\ell$. In this
way the non-linear interactions, originally a vector convolution in the 3D
vector space are reduced to to a 1D summation in terms of the shell index
$n$.  This one-dimensional model is the magnetohydrodynamic analog of the
GOY \citep[Gledzer-Ohkitani-Yamada: ][]{yam88b} shell-model of fluid
turbulence.

One obtains the following equation, given in \citet{giu98}:
\begin{equation}
\label{eq:dtzsh}
\frac{\mathrm{d}Z_n^\pm}{\mathrm{d}t} = -k_n^2 (\nu^+ Z_n^\pm + \nu^-
Z_n^\mp) + i k_n T_n^{\pm *} + f_n^\pm
\end{equation}
where $Z_n^\pm=u_n \pm b_n$ are the Els\"asser-like variables, $\nu^\pm =
(\nu \pm \eta)/2$ are combinations of kinematic viscosity and resistivity,
$f_n^\pm$ are external driving forces, and $T_n^\pm$ is the non-linear term,
obtained by assuming (1) that the non-linear interactions occur in triads of
neighboring modes and (2) the conservations of the total pseudo-energies
$E^\pm=\sum_n |Z_n^\pm|^2$ (and thus the energy $E=E^+ + E^-$ and the
cross-helicity $h_C=E^+ - E^-$) and a third invariant $H_K^{\alpha} = \sum_n
\mathrm{sign}(\delta-1)^n k_n^\alpha |v_n|^2$ which depends on the
dimensionality of the MHD system to model \citep{giu98}.

\subsection{Specificities of the \shellatm\ model}

The ``classical'' GOY-like shell-model that we have just presented
corresponds to MHD, where the average magnetic field \Bo\ has not yet been
separated out; in the \shellatm\ model we present now in this paper, the
average magnetic field \Bo\ is separated out, by starting from the RMHD
equations (\ref{eq:rmhd1}-\ref{eq:rmhd3}). The new model we obtain
corresponds basically to a pile of planes coupled by Alfv\'en waves and
containing each a ``classical'' shell-model for 2D MHD (Fig.~\ref{fig:loop}
top).  This is similar to the loop model developed independently by
\citet{nig04}, but the stratification of the atmosphere that we introduce
allows to use this model in a large variety of cases of coronal loops or
other structures (although we do not use this specific feature in the runs
presented in this paper).  In the \shellatm\ model:
\begin{itemize}
\item The profile of the Alfv\'en speed $b_{\parallel}(z)$ along the mean
  field (\ie the atmospheric structuring of the plasma) is given via a
  density stratification $\rho_0(z)$, an areal expansion factor of the flux
  tube $A(z)$ and magnetic flux conservation.  The latter two effects imply
  that the width of the loop and corresponding wavenumber $k_0$ must also,
  in general, depend on $z$.

\item The Els\"asser variables $Z_n^\pm$ now depend on the position $z$ of the
  plane along the main axis of the simulation box, and the left hand-side of
  Eq.~(\ref{eq:dtzsh}) is replaced by the term $(\partial_t \pm b_{\parallel}
  \partial_z) Z_n^\pm \pm \frac14 Z_n^\pm \partial_z(\ln\rho) \pm \frac12
  Z_n^\mp \partial_z b_{\parallel}$, corresponding to the linear Alfv\'en
  wave propagation in a stratified static atmosphere (Eq.~\ref{eq:velli93}).
  As a result, the external driving forces $f_n^\pm$ are not needed anymore,
  as energy can simply be input as an incoming energy flux at the
  boundaries.

\item The non-linear interactions occur inside each plane, in 2D. In this
  case the third invariant of MHD is anastrophy, \ie the total square module
  of the magnetic potential ($H_K^{\alpha}$ with $\alpha=2$). The
  coefficients of the non-linear terms $T_n^\pm$ of the shell-model are then
  those of \citet{giu98} with parameters $\alpha = 2$ and $\delta > 1$ (\ie
  $\delta = 5/4$ and $\delta_m = -1/3$).
\end{itemize}

To summarize, the fields of the \shellatm\ model we introduce and use in
this paper are the complex variables $Z_n^\pm(z,t)$, which are the
Els\"asser-like fields $u_n(z,t) \pm b_n(z,t)$. $n$ is is the index of the
shell, corresponding to the perpendicular wavenumber $k_n(z) = k_0(z)
\lambda^n$, with $\lambda=2$; it can be any integer (positive or negative),
but for numerical computations it is convenient to assume that
$Z_n^\pm(z,t)$ is $0$ outside some domain $[\![ 0, n_{\perp}-1
]\!]$\footnote{The energy flux from the domain $[\![ 0, n_{\perp}-1 ]\!]$
  outwards is then zero, as can easily be seen using the equation of the
  spectral energy flux~(\ref{eq:kflux}) for $n=0$ and $n=n_{\perp}$.}. $z$
is the position on the axis supporting \Bo, in a domain $[0,L]$ discretized
over $n_z$ planes.  The equations of the model are:
\begin{eqnarray}
  (\partial_t \pm b_{\parallel} \partial_z) Z_n^\pm 
  \pm \frac14 Z_n^\pm \partial_z(\ln\rho)
  \pm \frac12 Z_n^\mp \partial_z b_{\parallel}
  &=& \nonumber\\
  \qquad -k_n^2 (\nu^+ Z_n^\pm + \nu^- Z_n^\mp) + i k_n
  T_n^{\pm *} && 
  \label{eq:dtz}
\end{eqnarray}
with $T_n^\pm$ given by \citealt{giu98} (with $\alpha=2$, $\delta = 5/4$
and $\delta_m = -1/3$).

\subsection{Quantities derived from the fields of the model}
\label{sec:quant}

As $Z_n^\pm(z,t)$ represents the Els\"asser field at perpendicular
wavenumbers included in the shell $S_n$ and at position $z$,
$|Z_n^\pm(z,t)|^2/4$ is the energy per unit mass at position $z$ at time $t$
in the modes included in shell $S_n$. If we assume that the modelled loop is
a cylinder of diameter $2\pi/k_0$ and that, after discretization in the $z$
direction, the separation between planes (\ie the thickness of each plane)
is \dz, then the cross-section of the loop by a plane is $A=\pi^3/k_0^2$ and
the volume associated to each plane is $V=A\dz$; with a mass density
$\rho_0$, the mass associated to each plane is $m=\rho_0A\dz$ and the energy
contained in the field $Z_n^\pm(z,t)$ is:
\begin{equation}
  \label{eq:enshell}
  E_n^\pm(z,t) = \frac14 m |Z_n^\pm(z,t)|^2 = \frac14 \rho_0
  \frac{\pi^3}{k_0^2} \dz |Z_n^\pm(z,t)|^2
\end{equation}

$E_n$ as a function of $n$ (for any field, position and time) will hereafter
be referred to as the ``shell energy spectrum''. To get a 1D perpendicular
spectrum (as those given by turbulence theories), we need in addition to
take into account the geometry of the shell $S_n$ in Fourier space: for a
shell-model representing 2D MHD, $S_n$ has an area $\mathcal{S}(S_n) = \pi
k_n^2 (\lambda^2-1)$. It follows that the 2D energy spectral density in the
shell is $E_n / \mathcal{S}(S_n)$ and that the 1D energy spectral density is
$2 E_n / (k_n (\lambda^2-1))$. Note that for this reason there is a
difference of $1$ between the slope of a power-law 1D perpendicular spectrum
\citep[\eg $-5/3$ for a spectrum as][]{k41a} and the slope of its ``shell
energy spectrum'' counterpart (\eg $-2/3$).

\subsection{Scales of quantities of the model and time scales}
\label{sec:scales}

The equations are rendered non-dimensional introducing characteristic units
of time, length and density.  For the coronal situation we choose
$10^7\unit{m}$ for the unit of length, $1\unit{s}$ for the unit of time and
$10^9\unit{kg}$ for the unit of mass.  Then the units of the other
quantities derive naturally from these basic units and are $10\unit{Mm\cdot
  s^{-1}}$ for velocity, $10^{-12}\unit{kg\cdot m^{-3}}$ for mass density,
$10^{14}\unit{m^2\cdot s^{-1}}$ for diffusivities, $10^{23}\unit{J}$ for
energies, and $10^{23}\unit{W}$ for powers.

The characteristic time scales for each of the terms of Eq.~(\ref{eq:dtz})
are deduced from their orders of magnitude:
\begin{equation}
  \label{eq:dttimeshell}
  \partial_t Z  \sim b_{\parallel} \frac{k_{\parallel}}{2\pi} Z \sim \bar\nu
  k_{\perp}^2 Z \sim k_{\perp} Z^2  
\end{equation}
where $Z$ is an order of magnitude of the fields $Z_n^\pm$ of the model at
wavenumbers $k_{\parallel}$ (parallel to \Bo) and $k_{\perp}=k_0\lambda^n$,
and where $\bar\nu$ represents either $\nu$ or $\eta$. We obtain:
\begin{itemize}
\item the Alfv\'en time $\tau_A = 2\pi / (b_{\parallel} k_{\parallel})$;
\item the characteristic time of dissipation $\tau_{\nu} = (\bar\nu
  k_{\perp}^2)^{-1}$;
\item the characteristic time of non-linear interactions $\tnl =
  (k_{\perp} Z(k_{\perp}))^{-1}$ in the planes.
\item the wave reflection time scale $t_R = 2 / \partial_z b_{\parallel}$.
\end{itemize}

The maximum Alfv\'en time \tamax\ is obtained for $k_{\parallel}=2\pi/L$ and
corresponds to the time needed by the wave to cross the simulation box. By
taking on the other hand the minimum of all the characteristic times in the
box (using $k_{\parallel}=2\pi/\dz$ to get \tamin,
$k_{\perp}=k_{n_{\perp}-1}$ to get \tnumin, and $\tnlmin = ((k_{\perp}
Z(k_{\perp}))_{\mathrm{max}})^{-1}$), we can estimate the time step needed
by the numerical schemes according to Courant-Friedrichs-Lewy (CFL)
conditions.

Other time scales also appear in the \shellatm\ model:
\begin{itemize}
\item the correlation time $t^*$ of the forcing, which depends on the
  precise form of the forcing (see Sect.~\ref{sec:caseloop} for the case of
  a coronal loop);
\item the turbulent cascade time scale $\tau_{\mathrm{cascade}}=\sum_n
  \tnl(k_{\perp}=k_n)$ where the sum is taken on the modes $n$ of the
  inertial range of the spectrum (see Sect.~\ref{sec:spectra}).
\end{itemize}

\begin{figure}[tp]
  \plotone{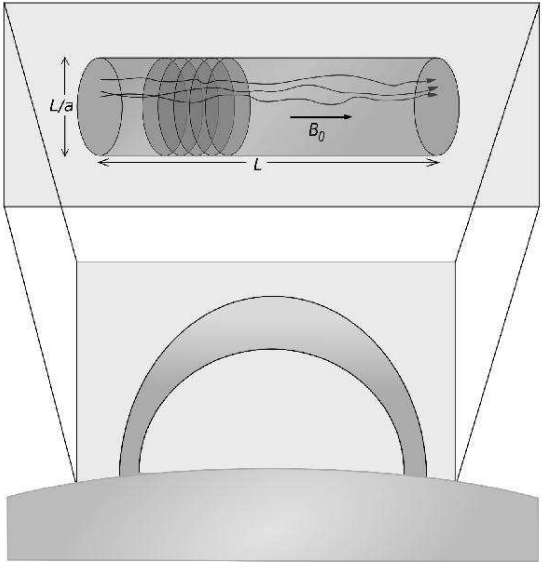}
  \caption{Top: layout of the \shellatm\ model in the general case;
    shell-models in planes orthogonal to \Bo\ are piled up along \Bo.
    Bottom: in the case of a coronal loop, the loop is unbent into the
    cylindric simulation box.}
  \label{fig:loop}
\end{figure}

\subsection{Case of a coronal loop}
\label{sec:caseloop}

\paragraph{Geometry} 

For the case of coronal loops forced via photospheric motions, we consider
the loops to be "straightened out" as seen in Fig.~\ref{fig:loop} (bottom).
This is similar to the cellular automaton model presented in \citet{buc03},
but here the non-linear interactions between modes of MHD turbulence are
represented through shell-models instead of cellular automata.  Furthermore,
for simplicity, we choose uniform density $\rho_0$, Alfv\'en speed
$b_{\parallel}$ and loop width ($2\pi/k_0$); more realistic cases will be
studied in future works.

\paragraph{Forcing} With this geometry, the boundary planes of the model
represent the footpoints of the loop, which are anchored in the photosphere.
We choose to impose a time-dependent vortex-like velocity field on modes at
the larger scales, corresponding to photospheric convective motions at the
scale of the supergranulation.  Since the velocity is imposed, waves
travelling along the loop are partially reflected at the photosphere.

The imposed velocity field $u_n(z,t)$ on each mode $n$ of both boundary
planes $z=0$ and $z=L$ has the form:
\begin{equation}
  \label{eq:shellforc}
  u_{z, n}(t) = u_{f,n} \left(
    \Exp{2 i \pi A_{z, n}} \sin^2 (\pi t / t^*) +
    \Exp{2 i \pi B_{z, n}} \sin^2 (\pi t / t^* + \pi / 2)
  \right)
\end{equation}
where $u_{f,n}$ is the amplitude of the forcing (non-zero only for some $n$
corresponding to scales $2\pi/k_n$ of the supergranulation), and $A_{z, n}$
and $B_{z, n}$ are each independent random complex coefficients of module
$1$ and whose complex argument is uniformly distributed over $[0,2\pi]$.
These coefficients are kept constant during time intervals of duration
$t^*$, and they are randomly changed when $t \equiv 0 \; [t^*]$ and $t
\equiv t^* / 2 \; [t^*]$ respectively, \ie when the corresponding $\sin^2$
term is zero.  This is another difference with the model of \citet{nig04},
who force using a stochastic velocity function on one boundary plane only.
The auto-correlation time of the forcing field is then of the order of
$t^*$, which is chosen to be much longer than all the other time scales of
the model (Sect.~\ref{sec:scales})

In practice, this boundary condition is realized by imposing an incoming
Alfv\'en wave $Z_n^+(0,t) = -Z_n^-(0,t) + 2 u_{0,n}(t)$ at the boundary $z =
0$ and an incoming Alfv\'en wave $Z_n^-(L,t) = -Z_n^+(L,t) + 2 u_{L,n}(t)$
at the boundary $z = L$.  The resulting power entering the loop is:
\begin{eqnarray}
  \epsilon_f &=& \frac14 \sum_n \rho_0(0) b_{\parallel}(0) A(0) 
    \left(|Z_n^+(0,t)|^2 - |Z_n^-(0,t)|^2\right)
    \nonumber \\ 
 && + \frac14 \sum_n \rho_0(L) b_{\parallel}(L) A(L) 
  \left(|Z_n^-(L,t)|^2 - |Z_n^+(L,t)|^2\right)
  \label{eq:incpow}
\end{eqnarray}
Note that this power is not imposed, but it depends on the fields already
contained in the simulation box.


\section{Numerical simulations and analysis of the results}
\label{sec:an}

\subsection{Energy}
\label{sec:energy}

\subsubsection{Energy balance}
\label{sec:enbal}

\begin{figure}[tp]
  \includegraphics[width=.85\linewidth]{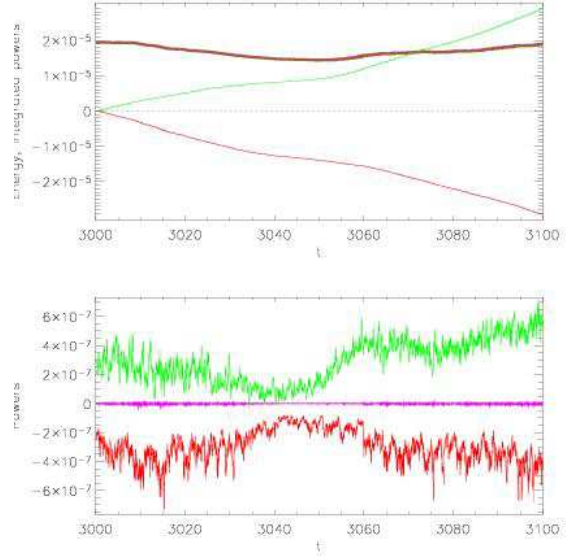}
  \caption{Energy balance in the model. Top: energies and integrated
    dissipation powers (top thin line: integrated power of forcing; bottom
    thin line: integrated dissipation power; thick line: energy, and sum of
    integrated contributions of powers).  The small deviation (only 1\,\%
    over the time span of this plot) of the energy compared to the sum of
    integrated powers shows that the numerical dissipation is low. Bottom:
    power time series (from top top bottom: forcing power, numerical
    dissipation power, dissipation power).  Quantities with negative
    contributions to the energy balance are shown as negative.  (color
    version online)}
  \label{fig:enbal}
\end{figure}

Alfv\'en wave propagation as well as the non-linear terms in our shell-model
conserve the energy, so that changes in total energy in the loop arise only
from flux through the photospheric boundaries (\ie the forcing) and from the
dissipation. This energy balance is well verified in practice, within $1\%$
in general as can be seen in Fig.~\ref{fig:enbal}, as long as the numerical
dissipation due to the numerical scheme for wave propagation is not too
high; the condition for this is that the perpendicular dissipation scales
are not too small compared to the separation between planes in the $z$
direction.

\subsubsection{Ratio of magnetic over kinetic energy}
\label{sec:mken}

The ratio of the magnetic to kinetic energy in the stationary state may be
estimated as follows.  First, a simple linear estimate of the velocity field
leads to:
\begin{equation}
  \vec u_{\perp} = \vec u_{\perp,0}(x,y) \cos (\omega_{ph}t)z/L,
  \label{eq:up2}
\end{equation}
while the magnetic field is given by:
\begin{equation}
  \vec b_{\perp} = \vec u_{\perp,0}(x,y) b_{\parallel} t/L.
  \label{eq:bp2}
\end{equation}

The relative importance of the higher frequency modes to this low-frequency
energy flux was discussed by \citet{mil97}: given that the Alfv\'en wave
travel times along loop is of the order of seconds, while most of the power
in photospheric motions is in the minutes to hour ranges, it is the
low-frequency resonance which plays a more important role.  Energy injection
from the photosphere into the corona therefore grows as $t^2$, and is stored
in the transverse coronal magnetic field, while the velocity field is
bounded by its photospheric value.  The linear solution will eventually
break down because the magnetic field determined in Eq.~(\ref{eq:bp2}) is
not in general force-free and therefore will cause the coronal field to
evolve dynamically.  The ratio of magnetic to kinetic energies at this point
may be estimated dimensionally by asking for the change in coronal velocity
field determined by nonlinear interactions in Eq.~(\ref{eq:rmhd1}) to be of
the same order of magnitude as the field given by Eq.~(\ref{eq:up2}).
Denoting the rms photospheric speed by $u_f$, after a time $\Delta t$ the
nonlinear term has the dimensional value $u_f^2 b_{\parallel}^2 \Delta t^2 /
\ell L^2$, growing quadratically with time. It will cause a change in the
coronal loop velocity field of the same order as $u_f$ over the time $\Delta
t$, when $u_f^2 b_{\parallel}^2 \Delta t^2 / \ell L^2 \sim u_f/\Delta t$.
One then recovers the time $\Delta t$ as $\Delta t \sim \tamax^{2/3}
\tau_e^{1/3}$ where $\tamax$ is the loop crossing time while $\tau_e = \ell
/ u_f$ is the non-linear time calculated on the photospheric velocity.  The
ratio of magnetic to kinetic energies in the corona at this stage is then
\begin{equation}
  \frac{E_b}{E_u} = 3 \left(\frac{\Delta t}{\tamax}\right)^2 =
  3 \left(\frac{\tau_e}{\tamax}\right)^{2/3}. \label{eq:enratio}
\end{equation}
which we associate with the saturation level of magnetic to kinetic energies.
This ratio should be a function of the aspect
ratio $a=b_{\parallel}\tamax/\ell$ of the loop. To check whether the
\shellatm\ model follows this dependence, we perform a series of simulations
with parameters $b_{\parallel} = 1$, $k_0=20\pi$ (\ie a width $\ell=0.1$),
$\nu=\eta=10^{-9}$, and $u_{f,n}=10^{-3}$ for $n\in [\![ 2,4 ]\!]$. The
number of shells is $n_{\perp}=16$ and the number of planes $n_z$ is taken
in the set $\{200, 500, 1000, 2000, 5000\}$ with a separation $10^{-3}$
between planes in all cases, leading to lengths $L \in \{0.2, 0.5, 1, 2,
5\}$ and aspect ratios $a \in \{2, 5, 10, 20, 50\}$.

\begin{figure}[tp]
  \plotone{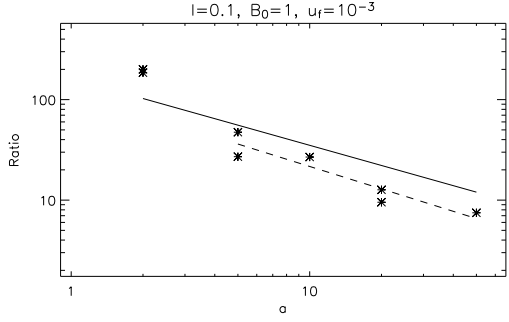}
  \caption{Ratio of average magnetic over kinetic energy as a function of
    aspect ratio, plotted with theoretical scaling (plain line) derived from
    Eq.~(\ref{eq:enratio}) and power-law fit (dashed line, with slope
    $-0.74\pm0.14$).}
  \label{fig:enratio}
\end{figure}

The ratio of magnetic energy to kinetic energy in the stationary state is
plotted as a function of aspect ratio in Fig.~\ref{fig:enratio}, together
with what is expected from Eq.~(\ref{eq:enratio}).  The numerical results we
obtain support roughly this scaling, although the experimental ratios are
smaller than the theoretical ratios by a factor $1.36$ and the values for an
aspect ratio $a=2$ deviate from the scaling obtained for other aspect
ratios.  The slight departure from the proposed scaling, and the
  fact that the saturation level of magnetic to kinetic energy is lower than
  predicted, could be due to the "leakage" of energy to the higher frequency
  resonances of the loop (as shown in Fig.~\ref{fig:tspec}).

\subsection{Spectra}
\label{sec:spectra}

\subsubsection{Formation of the spectra and spectral energy flux}

The energy flux in each plane from the shells $k < k_n$ to the shells $k \ge
k_n$ is the derivative of the energy contained in the shells $k \ge k_n$ due
to the non-linear terms $T_n^\pm$, namely:
\begin{eqnarray}
  \label{eq:kflux}
  \Pi_n &=& -\frac{k_n}{4\lambda^2} \Im \sum_{s = \pm 1}
  (\delta_m - \delta) Z_{n-2}^{-s} Z_{n-1}^s Z_n^s \nonumber \\
  && \qquad\qquad\quad +(2 - \delta - \delta_m) Z_{n-2}^s Z_{n-1}^{-s} Z_n^s
  \nonumber \\
  && \qquad\qquad\quad + \lambda \left(
    (\delta + \delta_m) Z_{n-1}^s Z_n^s Z_{n+1}^{-s} \right. \nonumber\\
  && \qquad\qquad\quad\quad \left. + (2 - \delta - \delta_m) Z_{n-1}^s Z_n^{-s}
  Z_{n+1}^s \right)
\end{eqnarray}
With $b_n=0$ we recover the hydrodynamic spectral energy flux given in
Eq.~(2) of \citet{fri98}.
Furthermore, this energy flux is consistent with the general idea that the
energy flows ``downhill'' in the 1D perpendicular energy spectrum.

\begin{figure*}[ht]
  \plotone{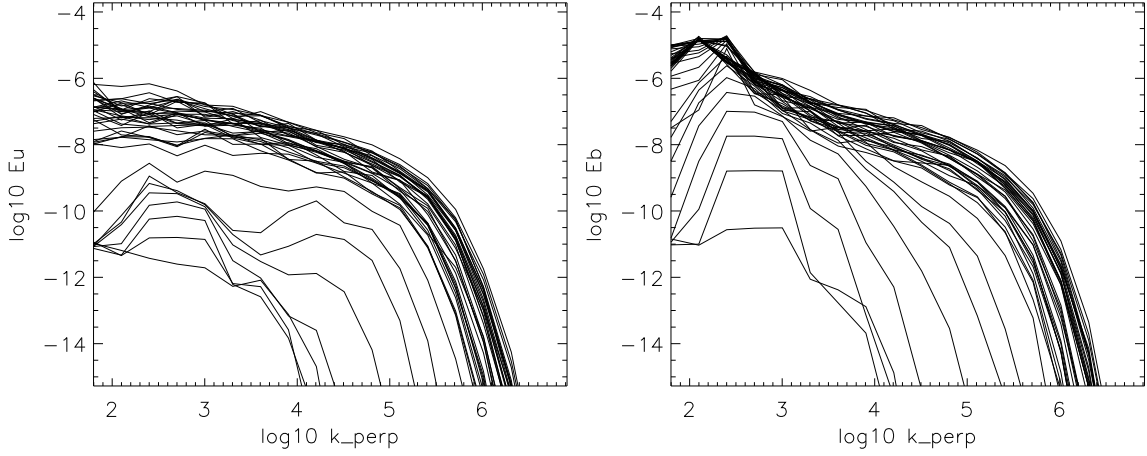}
  \caption{Kinetic (left) and magnetic (right) perpendicular spectra of
    energy in the shells of the model, averaged over the length of the
    loop. On each plot, 40 spectra are shown, separated by $10^{-2}$ units
    of time, starting shortly after the beginning of the simulation (lowest
    curves). The forcing is performed on modes corresponding to $\log_{10}
    k_n = 2.4$ to $3.0$. }
  \label{fig:specgrowth}
\end{figure*}

When starting the simulation from a very low amplitude field
(Fig.~\ref{fig:specgrowth}), the magnetic energy and then the kinetic energy
at the scales of the forcing grow, and when the fields are sufficiently
large, non-linear effects become visible as energy is transferred to modes
beyond those initially forced.  In particular, there is a flux to higher
$k_{\perp}$ mode numbers ($\Pi_n>0$, direct cascade), which continues to
highest wavenumbers where dissipation occurs, as well as a flux to smaller
wavenumbers ($\Pi_n<0$, inverse cascade) which energizes modes at the
largest scales and saturates at a level comparable to the forced modes.

As a result of the energy cascade, an inertial range appears between the
forcing and the dissipation scales, in the same way than in the original
shell-models.  The energy flux $\Pi_n$ across shells is uniform on average
over the whole inertial range. The Reynolds number in the case shown here
can be evaluated to $10^6$, which is much higher than any Reynolds number of
direct numerical simulations. Even higher Reynolds numbers can be attained
using more shells and planes, at the cost of the ability to do long-term
statistics.

\subsubsection{Fluctuations of the spectra}
\label{sec:spfluct}

\begin{figure}[tp]
  \plotone{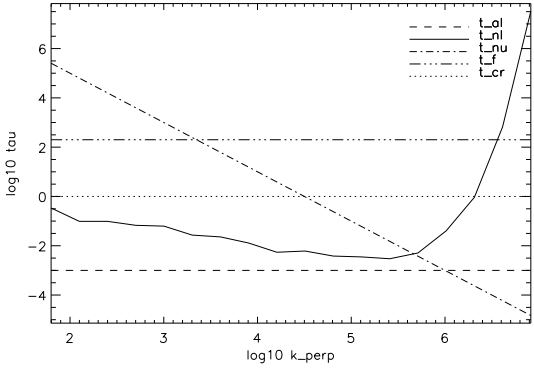}
  \caption{Time scales as a function of $k_{\perp}$: Alfv\'en time \tamin,
    non-linear time \tnl, dissipation time \tnu, forcing correlation time
    $t^*$, crossing time \tamax\ (from top to bottom in the caption inset).}
  \label{fig:dtk}
\end{figure}

\begin{figure}[tp]
  \begin{minipage}[c]{.88\linewidth}
    \plottwo{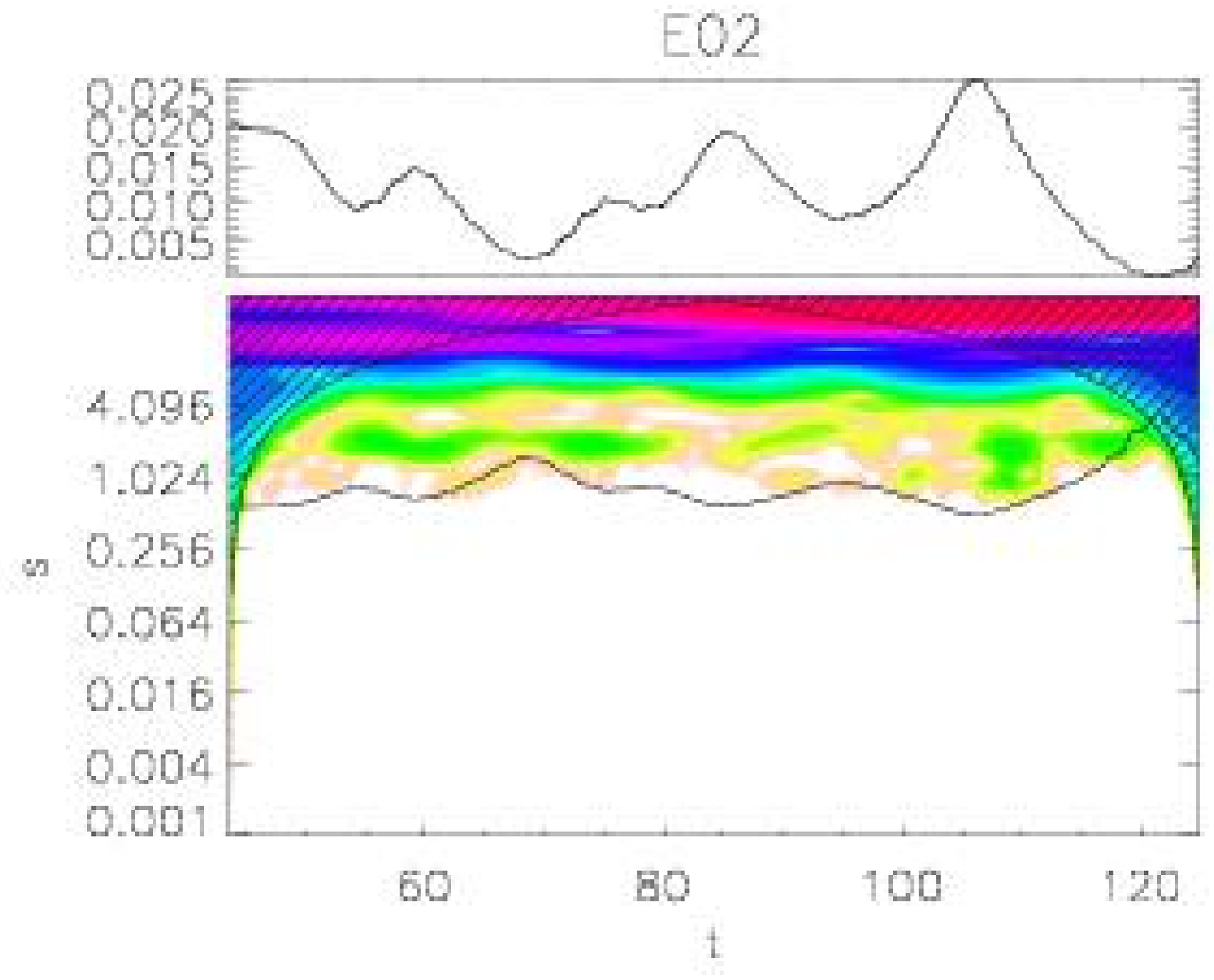}{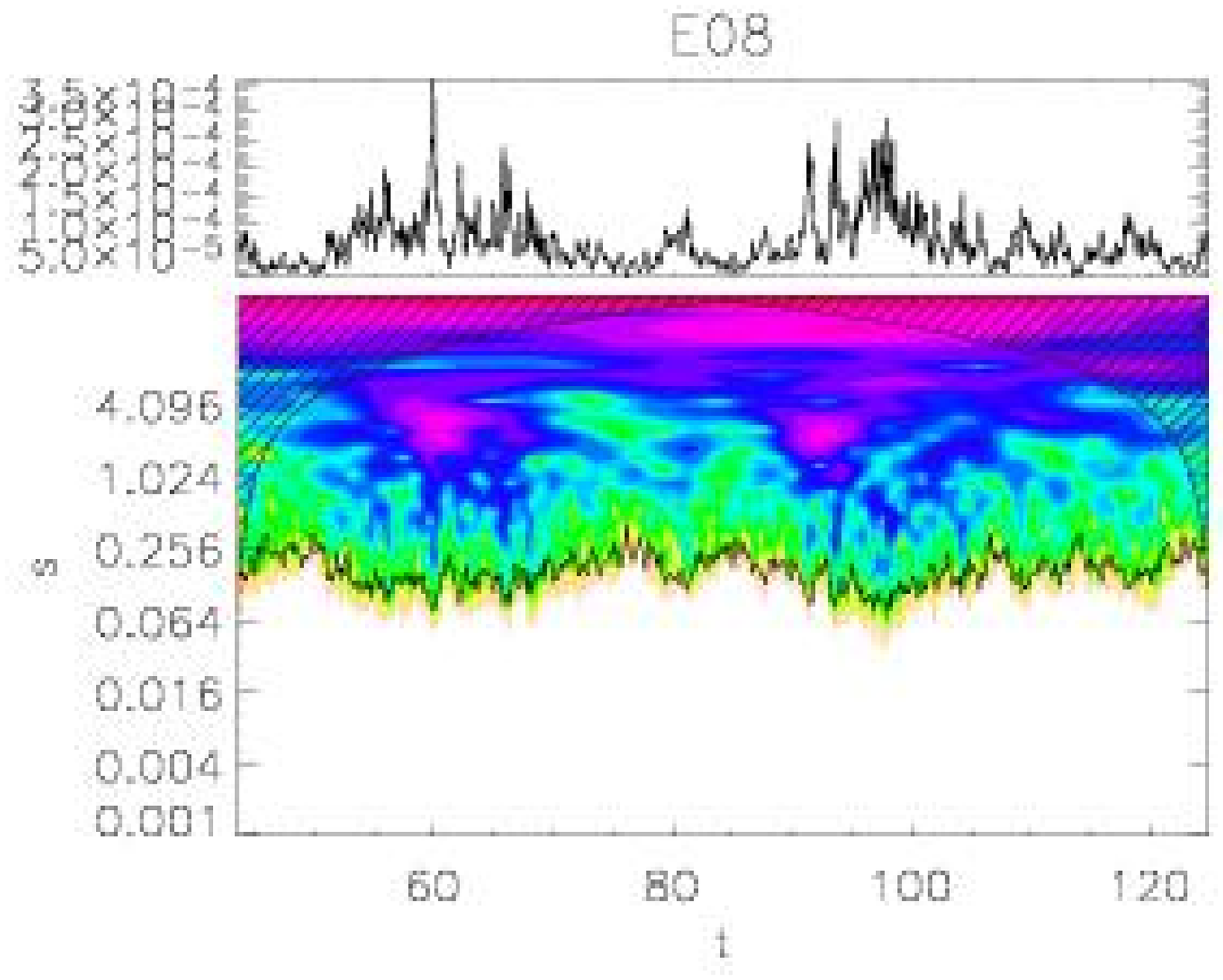}\\[2mm]
    \plottwo{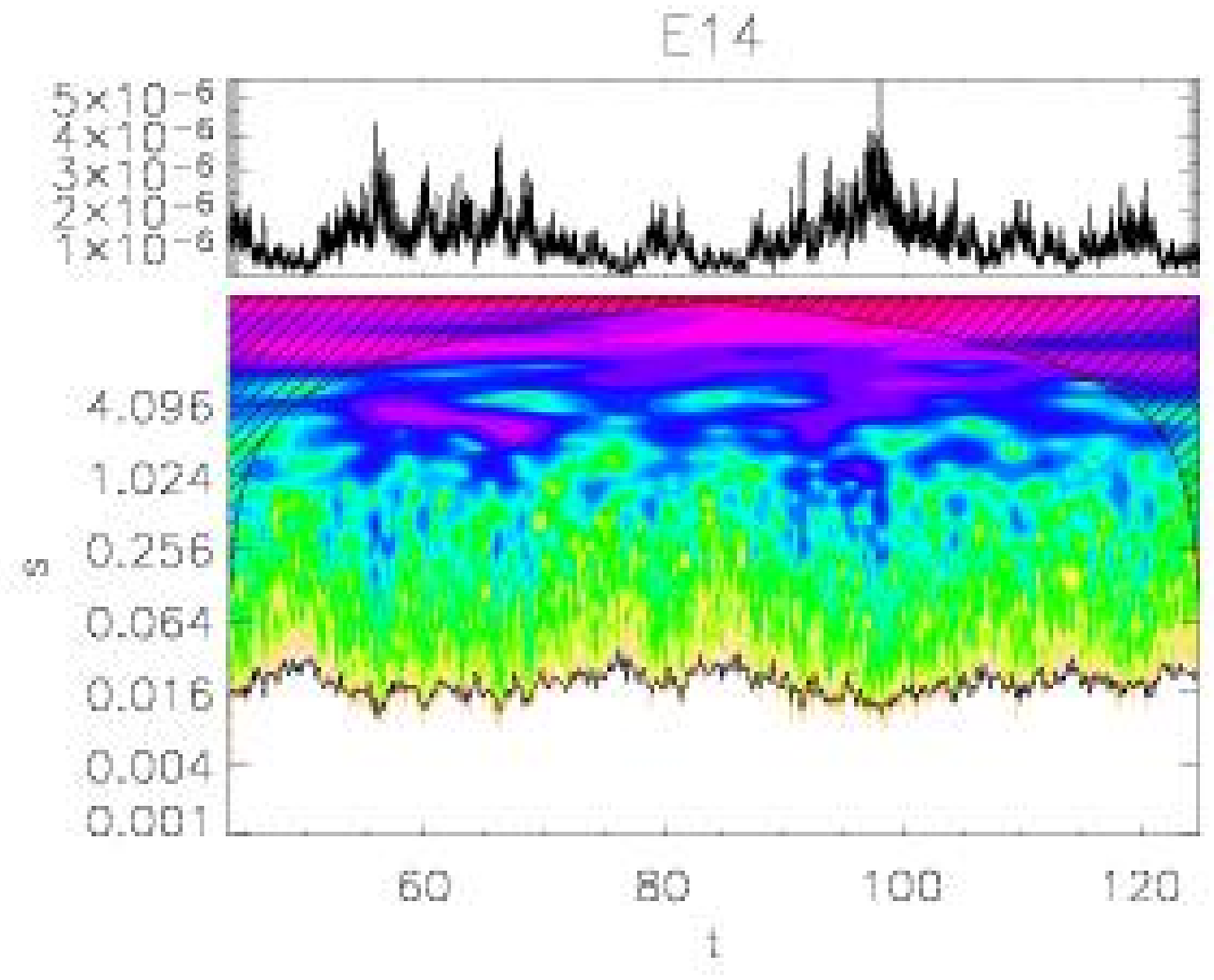}{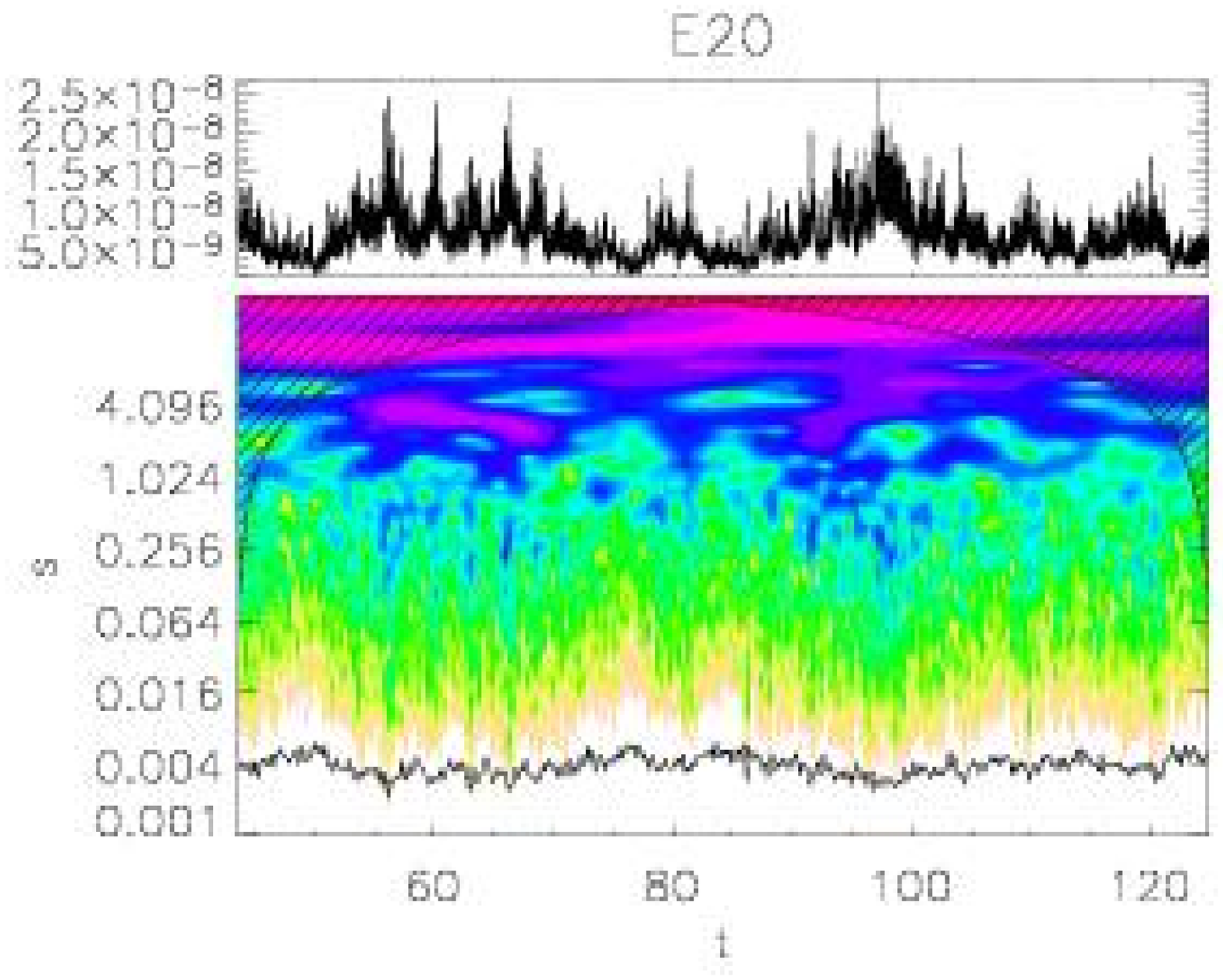}
  \end{minipage}%
  \begin{minipage}[c]{.09\linewidth}
    \includegraphics[width=\linewidth]{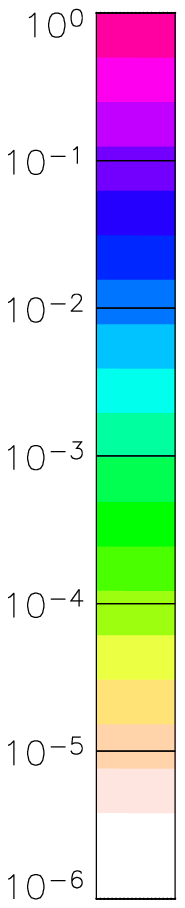}
  \end{minipage}
  \caption{Time series of the total energy contained in the shells
    $n\in\{2,8,14,20\}$ of the model, and Morlet wavelet time-scale planes
    of each of these time series. The axes of the time-scale planes are time
    (horizontally) and time scale (vertically, logarithmic). The theoretical
    non-linear time scale $\tnl(k_n)$ as a function of time has been
    superimposed on each time-scale plane. (color version online)}
  \label{fig:taunl}
\end{figure}

Once in the stationary phase, the spectra continue to fluctuate, with
characteristic time scales linked to the ``local'' time scales, \ie the time
scales, as described in Sect.~\ref{sec:scales}, considered as depending on
the mode $k_n$ (Fig.~\ref{fig:dtk}). The most relevant time scale seems to
be the local non-linear time scale $\tnl(k)$, defined in \ref{sec:scales},
as this is the time scale on which the energy in a given mode $n$ can change
under the action of the non-linear terms of Eq.~(\ref{eq:dtz}). More
precisely, no dynamics occur at time scales below the local non-linear time
scales, as can be seen in Fig.~\ref{fig:taunl}: the modes with low
$k_{\perp}$ have thus only long fluctuation times, excited by the long time
scales of the forcing, while the modes with high $k_{\perp}$ fluctuate fast,
but still with long characteristic times due to the flow of energy coming
from the modes at low $k_{\perp}$ (these long-term fluctuations are common
to the whole spectrum).

\subsubsection{Evolution of the spectra during an event. }

\begin{figure}[tp]
  \centering
  \plotone{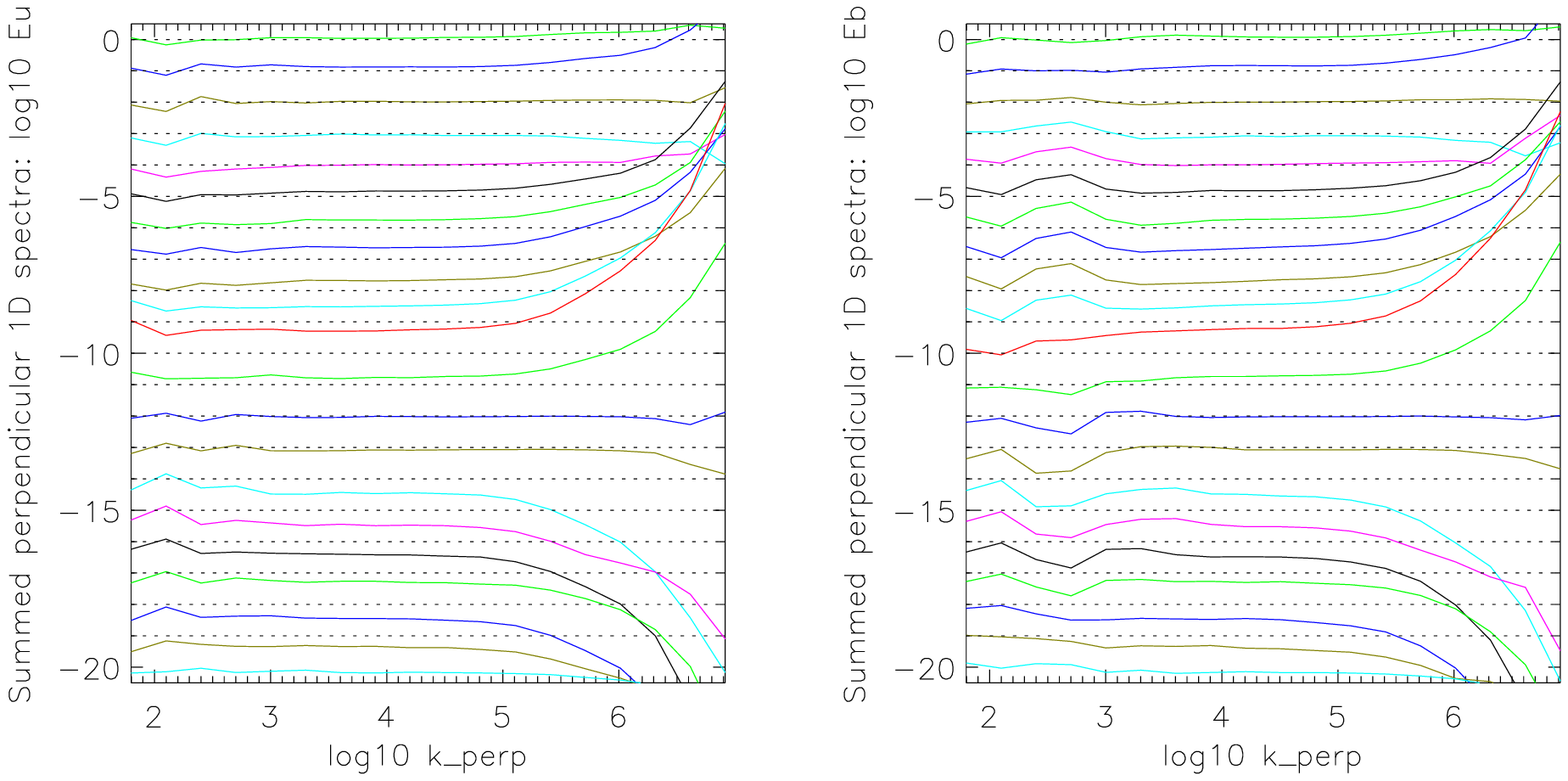}\\
  \plotone{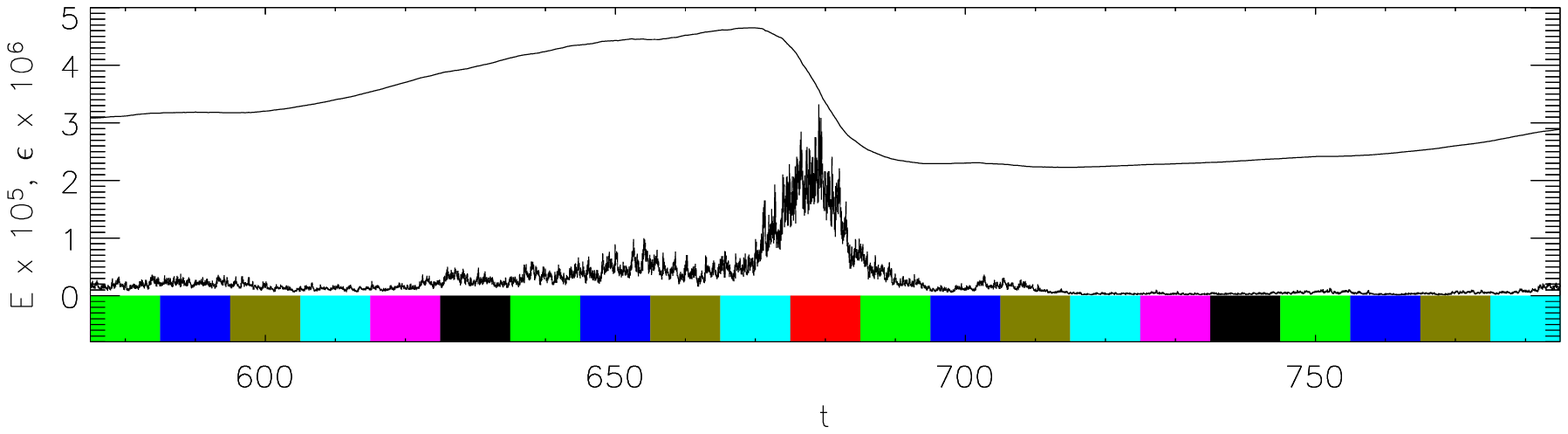}
  \caption{Top left: differences (in $\log_{10}$-space) between kinetic
    perpendicular spectra of energy in the shells of the model and their
    average (in $\log_{10}$-space).  The spectra are averaged over the
    length of the loop, and are plotted at times surrounding an event of
    dissipation power (corresponding to the red spectra): 10 spectra are
    shown before the event, and 10 spectra are shown after; the difference
    spectra are each separated by $10$ units of time, and are stacked from
    top to bottom, with a shift of 1 unit of the $y$-axis between each of
    them. Top right: same plot for the magnetic energy spectra. Bottom: time
    series of energy and dissipation power, with the colors corresponding to
    the time intervals used to compute the spectra.}
  \label{fig:specstack}
\end{figure}

To understand what happens during episodes of high energy dissipation, we
have analyzed the spectra of the fields before, during and after such an
episode. The differences of spectra in respect to an average spectrum
(Fig.~\ref{fig:specstack}) show that before the event (the maximum
dissipation corresponds to the red spectrum differences), the energy
accumulates over the whole spectrum.  The total energy is then high, the
non-linear times are short, and the energy flows down rapidly to the
smallest scales according to Eq.~(\ref{eq:kflux}): it enhances the
spectra at the largest wavenumbers by several orders of magnitude, leading
to a strong enhancement of the dissipation power. As energy is released,
this process leads then to a decrease of the spectrum, first in the
dissipative range (high wavenumbers), then in the whole spectrum. The
dissipation power is then low again, and as the non-linear time scales in
the inertial range are longer, the energy injected at the largest scales
cannot flow to the smallest scales as fast as before: the energy does not
reach easily the dissipative scales and the dissipation power remains low,
until the next such episode.

\citet{nig05} underline that the leading term of the energy flux across
scales (Eq.~\ref{eq:kflux}) is proportionnal to $k_nb_n^2u_n$ (with the
notations of the shell-model variables) and they also observe short-term
variations of the kinetic energy spectrum around a dissipation event.  These
variations appear to control the energy flux to the smallest scales, and
then the dissipation.  In addition, we have shown that these variations
exist on a longer term around an event, and that the magnetic energy
spectrum also varies on the same time scales.  The cross-scale energy flux
may thus be controlled by both the kinetic and the magnetic energy spectra.

\subsubsection{Slopes of the spectra}
\label{sec:spslope}

\begin{figure}[tp]
  \plotone{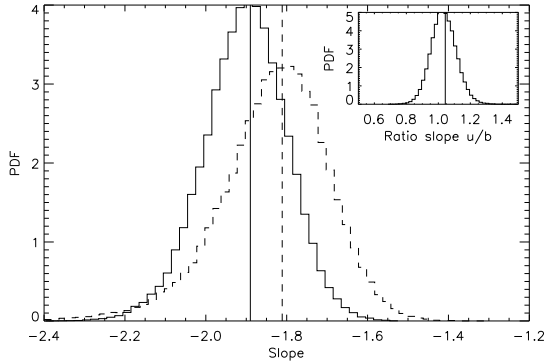}
  \caption{Distribution function of the slopes of the 1D perpendicular
    spectrum averaged along \Bo, for the kinetic (plain line) and magnetic
    (dashed) fields. The median slopes are respectively $-1.89$ and $-1.81$,
    and are plotted as vertical lines. The distribution of the ratios
    between the slopes for kinetic and magnetic perpendicular spectra is
    shown in the inset, together with its median $1.044$.}
  \label{fig:specslope}
\end{figure}

The slopes of the power-law 1D perpendicular spectra of the velocity and
magnetic field (Fig.~\ref{fig:specgrowth}) seem to be roughly equal on the
inertial range, but as the spectra fluctuate with time, there are
fluctuations of the slopes. The distribution of these slopes obtained at
different times is shown on Fig.~\ref{fig:specslope}: the median slope is
$-1.89$ (with a standard deviation $0.10$) for the velocity spectrum and
$-1.81$ (with a standard deviation $0.13$) for the magnetic spectrum. It
appears that, on average, the kinetic spectrum is slightly steeper (by
$4\;\%$) than the magnetic spectrum. If we look specifically at the times
when the total dissipation power exceeds its 90th percentile, \ie during
events of energy dissipation, the spectra are slightly shallower, with
medians of slopes being $-1.83$ and $-1.77$ for the velocity and magnetic
spectrum respectively. This translates the fact that more energy is present
at small scales during events of energy dissipation (red curves of
Fig.~\ref{fig:specstack}).

These 1D spectra are different than the ones found by \citet{nig05},
corresponding to 1D spectra of slopes $-5/3$ for velocity and $\approx -3$
for magnetic field; an explanation could be that their inertial range is
smaller, and that their fitting range includes scales where forcing occurs.

\subsubsection{Parallel and perpendicular spectra}
\label{sec:sppar}

In this model non-linear interactions occur only in perpendicular planes.
Development of small scales along the magnetic field is then merely a
consequence of the Alfv\'enic propagation of differences in the dynamics in
different planes.  One therefore expects parallel and perpendicular spectra
to be different, with a relationship determined by the so-called critical
balance condition, namely, that for a given perpendicular scale, differences
in the parallel direction can appear only between planes such that the
Alfv\'en propagation time is longer than the (perpendicular) non-linear time
at that same scale \citep[e.g.][]{gs95,cho02,oug04}. In the case of the
present model, assuming a $k^{-\alpha}$ 1D energy spectrum (\ie a
$k^{-\alpha+1}$ ``shell energy spectrum''), the non-linear time scale is
$\tnl(k_{\perp}) \propto k^{(\alpha-3)/2}$.  With a constant and uniform
advection velocity $b_{\parallel}$, the critical balance condition can be
expressed by:
\begin{equation}
  \label{eq:critbal}
  k_{\parallel} \la \frac{Z(k_0)}{b_{\parallel}} k_{\perp}^{\frac{3-\alpha}2}
  k_0^{\frac{1-\alpha}2}
\end{equation}
Note that with a Kolmogorov $\alpha=5/3$ spectrum, we recover the result
$k_{\parallel} \propto k_{\perp} ^{2/3}$ of \citet{gs95}.

For a field $a_n(z)$ of the model at a given time $t$ ($a$ can be $Z_n^\pm$,
$u_n$ or $b_n$), let $\tilde a_n(k_{\parallel})$ be its Fourier transform
along the $z$-axis. We obtain the two-dimensional power spectrum of $a$
(function of $k_{\perp} = k_n$ and $k_{\parallel}$) from:
\begin{equation}
  \label{eq:2dspec}
  \mathcal A(k_{\perp},k_{\parallel}) = \frac{c}{k_n} \left|\tilde
    a_n(k_{\parallel})\right|^2
\end{equation}
where $c$ is a constant.

\begin{figure}[tp]
  \plotone{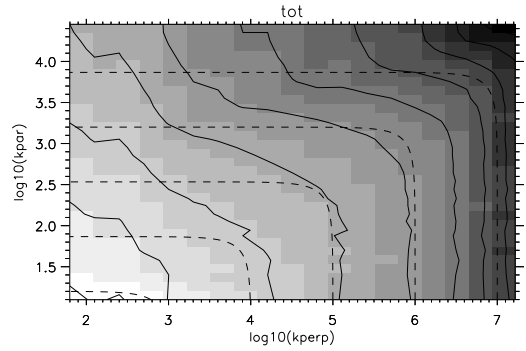}
  \caption{Total spectrum of the $Z^+$ and $Z^-$ fields, as a function of
    the perpendicular and of the parallel wavenumbers. The plain lines are
    level lines, and the dashed lines are ellipses of axes $k_{\perp}$ and
    $k_{\parallel} \propto k_{\perp}^{2/3}$, for different values of
    $k_{\perp}$.}
  \label{fig:spec2d}
\end{figure}

To get a sufficient wavenumber range in the parallel and perpendicular
directions, we need to do simulations with a very large number of planes.
This is achieved by starting a simulation with a number of planes $n_{z,0}$,
and then, once the energy has reached its final order of magnitude, by
stopping the simulation and resuming it after having interpolated the fields
in the $z$ direction. We can perform several steps of this process if
needed.  Figure~\ref{fig:spec2d} shows a 2D spectrum obtained by summing the
$Z^+$ and $Z^-$ spectra and by averaging them over 10 times separated by
$10\tamax$, during a run with $n_z=10n_{z,0} = 5000$ planes (runs with
$50\,000$ planes were also performed).  The level lines in the
$(k_{\perp},k_{\parallel})$ space are clearly non-circular, and appear to
follow the critical balance ellipses (Eq.~\ref{eq:critbal}) at large
$k_{\perp}$ though with an excess of energy in the parallel direction.  The
anisotropy angle as defined by Eq.~(5) of \citet{delz01}, computed on the
range $\log_{10}k\in [1.8, 4.4]$ (where the spectrum is known as a function
of both $k_{\perp}$ and $k_{\parallel}$) is 67 degrees, which confirms that
the spectrum is elongated in the perpendicular direction.

\subsection{Dissipation,  heating function and statistical properties}
\label{sec:diss}

\begin{figure}[tp]
  \includegraphics[width=\linewidth]{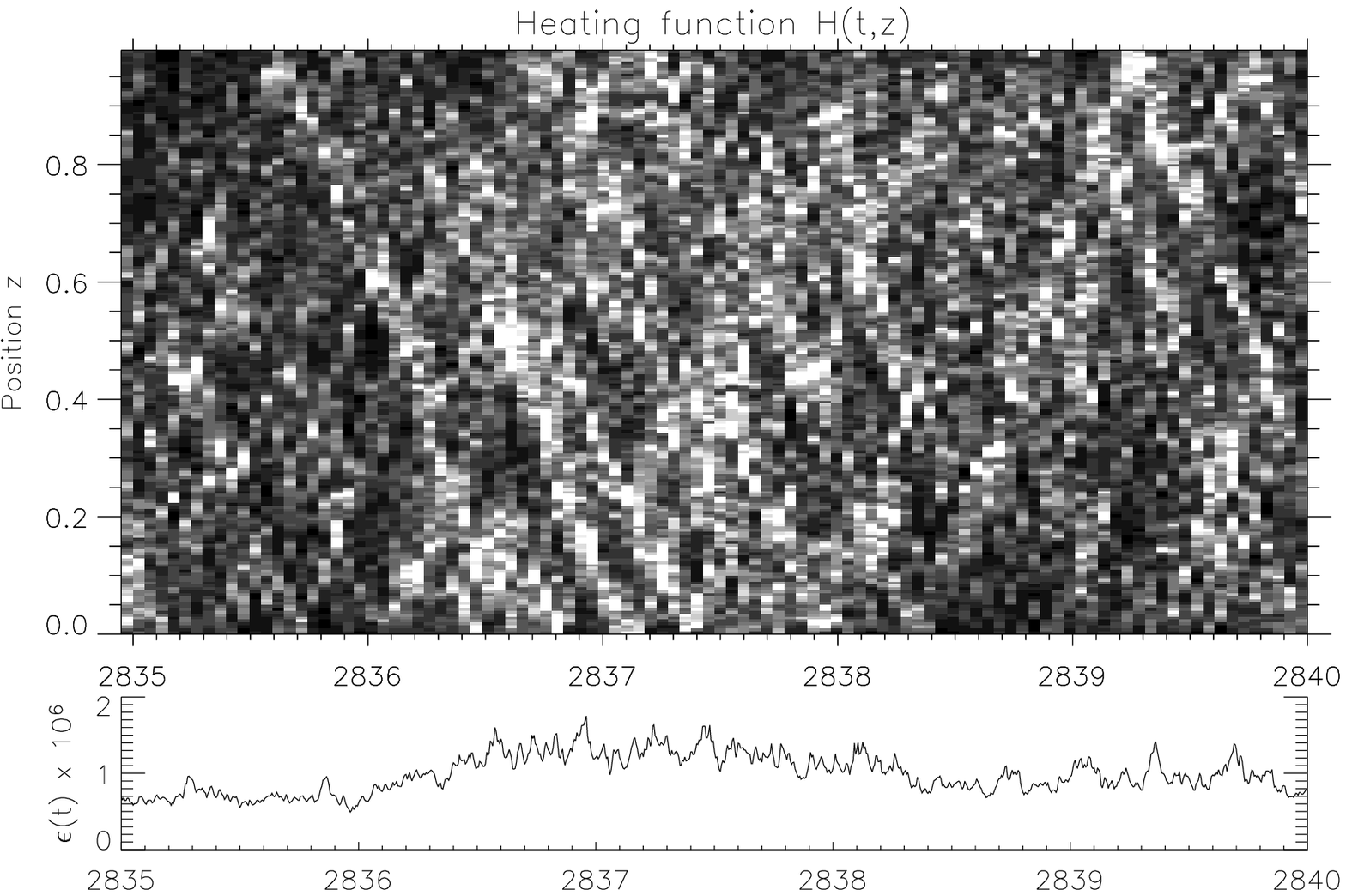}\\[3mm]
  \includegraphics[width=\linewidth]{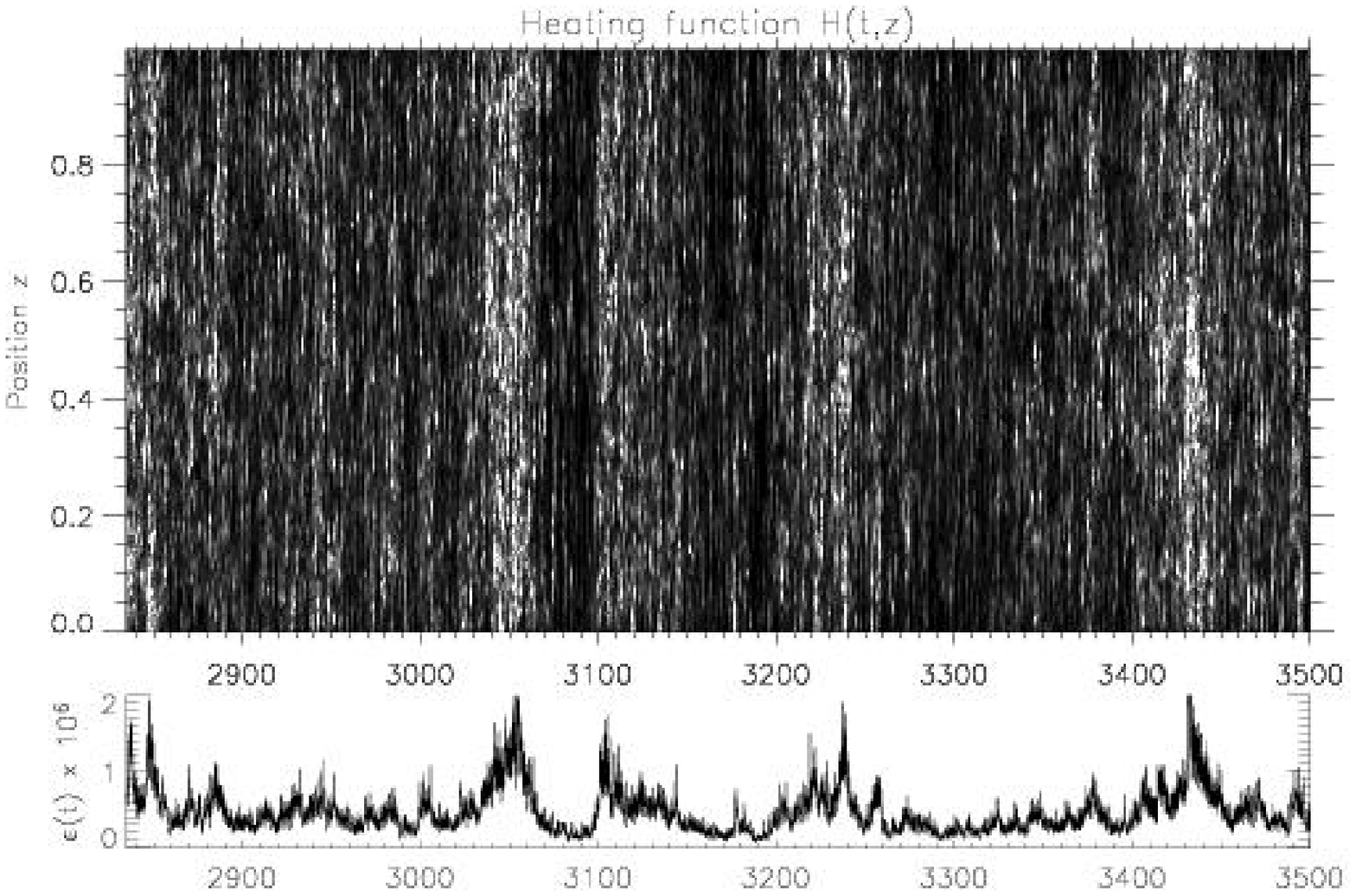}\\[3mm]
  \includegraphics[width=\linewidth]{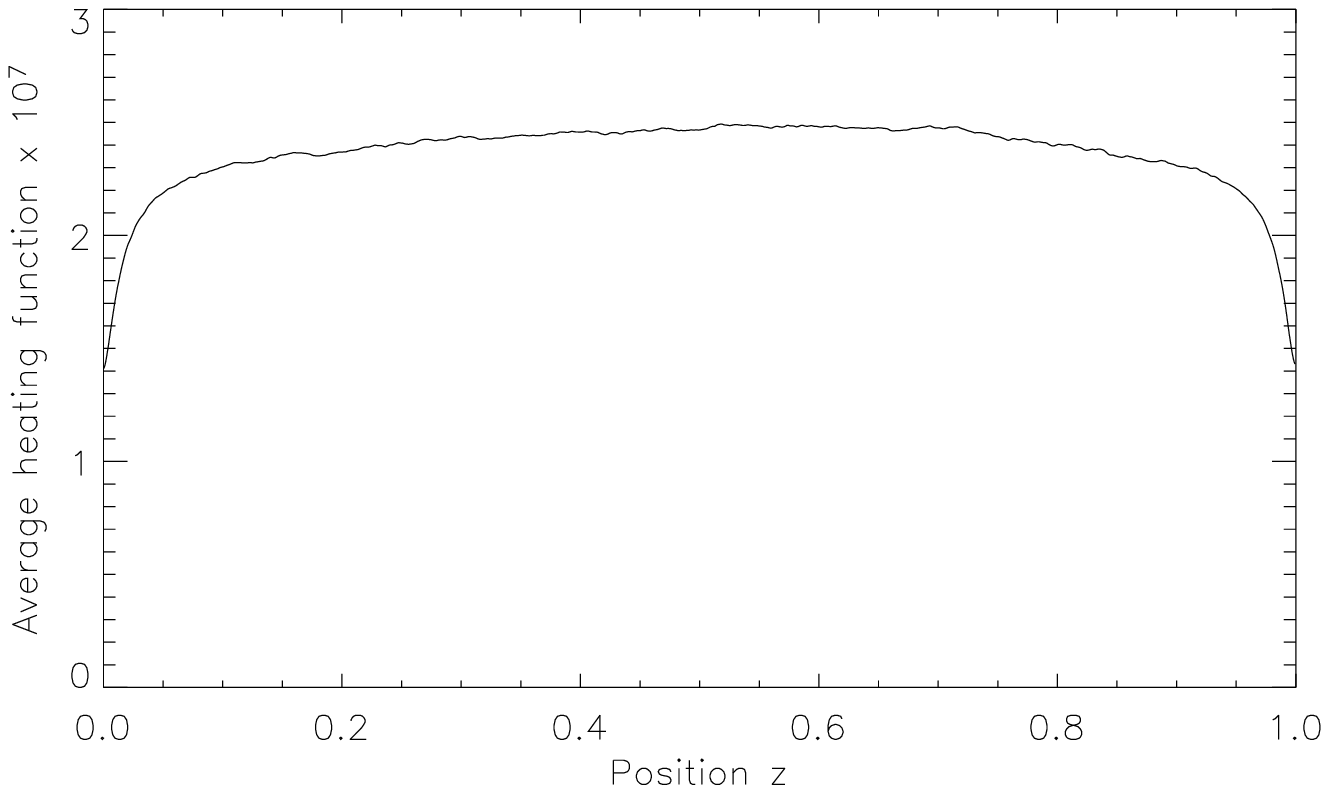}
  \caption{(Top and middle panels) Heating function, or power of energy
    dissipation per unit length as a function of time $t$ and position $z$
    along the loop. Two different time intervals are shown in sub-figures
    (the top and middle panels). The bottom panels in each sub-figure
    represents the integral along the loop of the heating function (\ie the
    total power of energy dissipation as a function of time). (Bottom panel)
    Time average, over $1200\tamax$ following (top panel), of the heating
    function as a function of the position along the loop.}
  \label{fig:heatfunc}
\end{figure}

\paragraph{Heating function.} If we look at the energy dissipation power per
unit length as a function of both time and position along the loop, we get
the ``heating function'' (Fig.~\ref{fig:heatfunc}a). We see again (and for
the same reasons than before) short-lived events of dissipations, and they
correspond to short structures along the axis of the loop, whose size is of
the order of the propagation distance of the structure during its lifetime.
Some Alfv\'en wavepackets are also strong enough to be dissipated only after
interacting with a lot of counter-propagating wavepackets, and thus they
live a longer time and leave an oblique trace in the heating function during
their propagation.

Furthermore, when we look at the heating function at long time scales of
hundreds of crossing times \tamax\ (Fig.~\ref{fig:heatfunc}b), some features
appear, which are related to the slow variations of the total energy (mainly
contained in the slow-variating low-$k_{\perp}$ modes) under the effect of
the slow forcing of time scale $t^*$ (which is chosen to be a few hundreds
of \tamax). The time variations of the dissipation power at these times
scales (corresponding to a few minutes of physical time) seem to be almost
the same on all positions along the loop. This is consistent with the common
statement that the loop is heated as a whole, even though (1) the elementary
events of dissipation, as seen in Fig.~\ref{fig:heatfunc}a, are each small
compared to the length of the loop (2) thermodynamics, which would further
smooth out the appearance of the heating function obtained from observable
variables (because of the fast conduction times) has not yet been taken into
account.

\paragraph{Dissipation power time series.} The integral of the heating
function along the loop is the time series of the power of energy
dissipation $\epsilon(t)$, shown in the bottom panels of
Figs.~\ref{fig:heatfunc}(a-b). These time series display spikes of high
dissipation power at short time scales during high-activity periods, as is
usually found in both observations of solar flares and simulations of
high-Reynolds number MHD turbulence. This dissipation time series will be
analyzed further later.

\paragraph{Average profile of dissipation power.} On the other hand, the
time average of the heating function, \ie the average power of energy
dissipation per unit length as a function of the position along the loop,
shown in Fig.~\ref{fig:heatfunc}(c), is almost flat and drops only near the
loop footpoints. This would suggest that coronal heating takes place almost
uniformly along loops, although not at footpoints; however one must bear in
mind that these simulations do not yet take into account realistic profiles
of density and magnetic field.

\begin{figure}[tp]
  \plotone{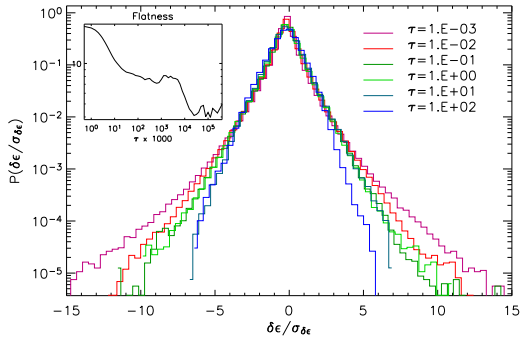}
  \caption{Distributions of the increments of the energy dissipation power
    time series, for different time lags. Inset: flatness corresponding to
    these dissipation power increments. (color version online)}
  \label{fig:interm}
\end{figure}

\paragraph{Intermittency}
The increments $\delta_{\tau}\epsilon(t) = \epsilon(t+\tau) - \epsilon(t)$
of the time series $\epsilon(t)$ at a given time lag $\tau$ have a
distribution whose shape depends on the time lag $\tau$: in
Fig.~\ref{fig:interm}, the distributions of the $\delta_{\tau}\epsilon(t)$
normalized by their standard deviation have wider wings for short time lags
than for long time lags.  Hence the time series is intermittent, which is
confirmed by the rise of the flatness (fourth normalized structure function)
$F(\tau)=\langle |\delta_{\tau}\epsilon(t)|^4 \rangle_t / \langle
|\delta_{\tau}\epsilon(t)|^2 \rangle_t^2$ for small time lags $\tau$ (inset
of Fig.~\ref{fig:interm}).  This intermittency is a consequence of the
intermittency that can be observed in the velocity and magnetic fields of
the model, and which is also predicted by models such as \citet{shel94} in
hydrodynamics and \citet{polit95} in MHD.  It could be a consequence of the
fluctuations of the spectral flux resulting from the long-term global
fluctuations of the spectrum which have been seen in
Sect.~\ref{sec:spfluct}.  The modes with high $k_{\perp}$ are then
intermittent, and as they participate the most to energy dissipation, the
time series of energy dissipation power is intermittent.

\begin{figure}[tp]
  \plotone{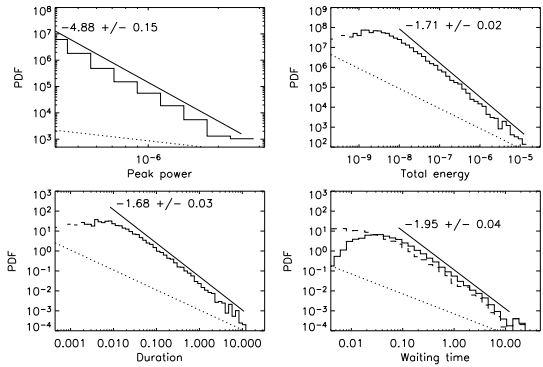}
  \caption{From left to right and from top to bottom: distributions of peak
    power, total energy content, duration, and waiting times, for events
    found in the time series of energy dissipation power.  The dashed line
    represents one event per histogram bar. }
  \label{fig:events}
\end{figure}

\paragraph{Events}
Statistics issued from observations often involve the detection of events,
or structures, from the observed fields \citep{asc00,parn00,buc06}, and the
distributions of their characteristics. Following the ``threshold''
definition of \citet{buc05}, with a threshold fixed at the average
dissipation power, we get the distributions shown in Fig.~\ref{fig:events}
for the event total energy content, the peak power of energy dissipation,
the duration of events, and the waiting-time between two successive events.

The distribution of the peak power in events is narrow, as a result of the
summation of the heating function over the whole loop: the local spikes of
energy dissipation are hidden by the average dissipation occurring in the
whole loop. On the other hand the distributions of integrated dissipation
power (total energy content of events) and of event durations are very wide.
This is partially due to the threshold definition used \citep{buc05}, in the
case of this time series where long time scales are superimposed to the
shorter time scales of energy dissipation in the dissipation range.
Furthermore, the waiting-times between successive events have also a wide
power-law distribution. However, as discussed extensively in \citet{buc05},
this result depends on what definition of an event is used to extract events
from the time series of the power of energy dissipation.

Compared to the distributions of events obtained from the loop shell-model
of \citet{nig04}, the main difference is the much steeper slope ($-4.9$
instead of $-1.8$) of the distribution of the peak power in events.  The
reason could be the summation effect due to the existence of more but
smaller dissipation events along the loop, because of the higher resolution
we used in this run, both along the loop ($n_z=2000$ instead of $n_z=200$,
allowed by the parallelization of our code) and in the perpendicular
direction ($n_{\perp}=16$ instead of $11$).

\subsection{Frequencies and time correlations}
\label{sec:tcorr}

\subsubsection{Frequencies}
\label{sec:freq}

\begin{figure}[tp]
  \includegraphics[width=\linewidth]{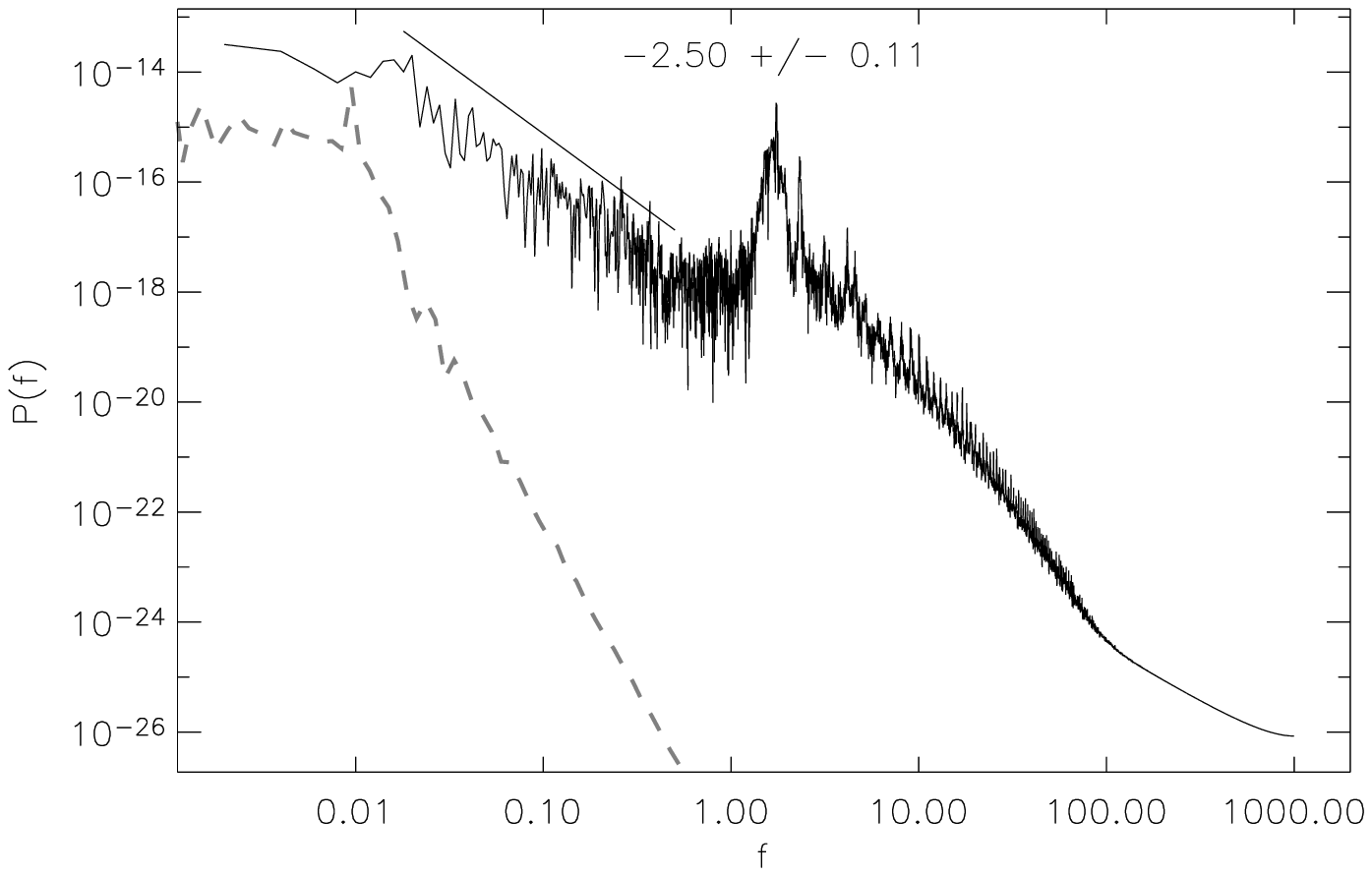}
  \includegraphics[width=\linewidth]{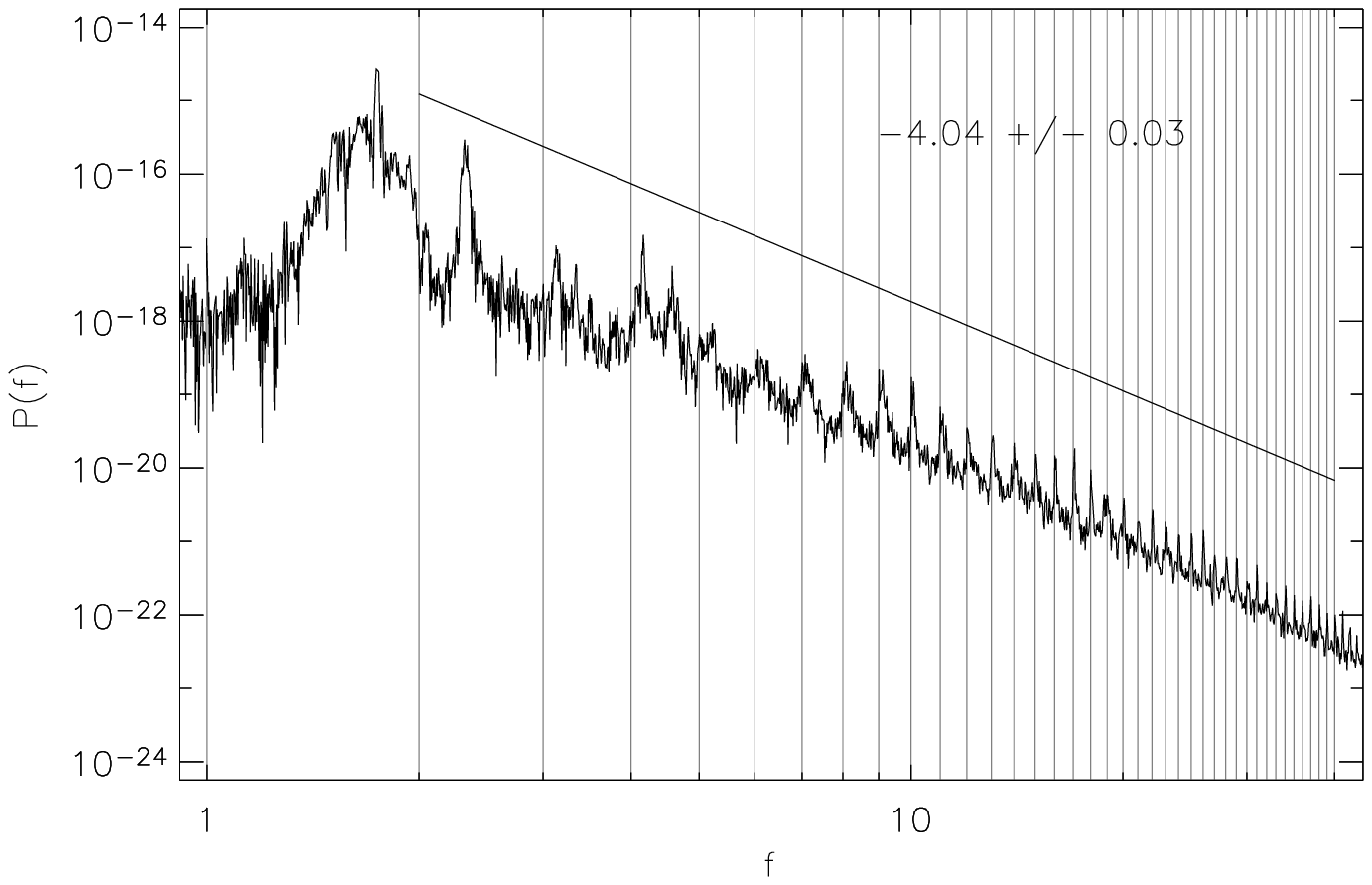}
  \caption{Plain black lines: spectrum of the time series of kinetic energy.
    The bottom panel is a zoom on the high-frequency range of the spectrum,
    where vertical lines represent the harmonics of the first resonance
    frequency $f_0=1$.  Power-law fits and the slopes obtained are
    superimposed; the horizontal range of the lines indicates the range of
    the fits, and they are shifted vertically for clarity.  The spectrum (in
    arbitrary units) of the square amplitude of the forcing function
    ($|u_{0,n}(t)|^2$ from Eq.~\ref{eq:shellforc}) is superimposed on the
    top panel (dashed grey line). }
  \label{fig:tspec}
\end{figure}

The time series of the kinetic and magnetic energies reveal oscillations
corresponding to exchange of energy between the velocity and magnetic field.
These exchanges occur, for example, thanks to the crossing of $Z^+$ and
$Z^-$ wavepackets, and should have periods which vary depending on the
precise number and repartition of wavepackets along the loop.  A
characteristic time scale is of course given by the Alfv\'en crossing time
\tamax, corresponding to the first resonance frequency $f_0 = 1 / \tamax$.
Multiples of the Alfv\'en crossing time correspond to higher-frequency
resonances of a loop in linear theory.  While the power spectrum of the time
series of total energy is a power-law of index $-2$ over more than $4$
decades, the spectra of magnetic or kinetic energy display peaks
corresponding to these resonances. The spectrum of the time series of
kinetic energy (Fig.~\ref{fig:tspec}) fits to a power-law of index $-2.5$ at
very low frequencies.  The first resonant frequency, together with the
higher frequency harmonics, appear as peaks overlying a different, steeper
power-law for the higher frequencies, shown in the bottom panel of
Fig.~\ref{fig:tspec}.  The frequencies of the peaks correspond well with the
integer multiples $nf_0$ of the fundamental for $n \geq 5$, while at lower
frequencies they appear shifted.  This shift, which is absent in a linear
simulation (realized with the same parameters but without shell-models, \ie
with no non-linear interactions), is probably due to anharmonicity
introduced by the non-linear effects, as shown by \citet{mil97} or
\citet{nigphd}.

\begin{figure}[tp]
  \plotone{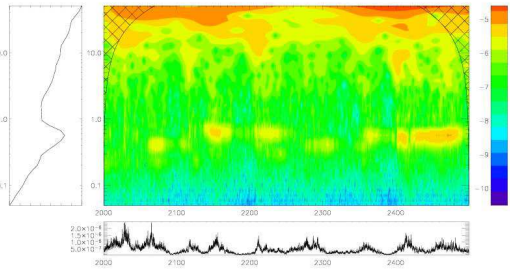}
  \caption{Morlet wavelet time-scale plane of the kinetic energy time series
    (logarithmic color scale). Oscillations of long but finite lifetime and
    of different frequencies can be seen at time scales (vertical axis)
    between $0.2$ and $0.6$. Left panel: average wavelet spectrum. Bottom
    panel: time series of total dissipation power. (color version online)}
  \label{fig:wspec}
\end{figure}

An even better understanding of these oscillations may be gleaned from a
time-frequency analysis via wavelet transform shown in Fig.~\ref{fig:wspec}:
there are oscillations which have long but finite lifetimes and different
frequencies dominantly around the fundamental harmonic.  These oscillating
high-frequency wavepackets appear to arise in association with dissipation
bursts, seen in the dissipation power time-series (bottom panel of
Fig.~\ref{fig:wspec}).  This intermittent rise in the high-frequency
component of the velocity field may be involved in the enhanced non-linear
interactions required to generate the bursts in power, as required in the
flare-driving mechanism highlighted in \citet{nig05}.  On the other hand,
their persistence may be related to excitation by the time-space
localization of the burst themselves, a sort of post-microflare resonant
ringing, which might be observable with future high-cadence spectroscopic
measurements.

The comparison between the spectra of the forcing function at a boundary and
of the resulting energy time series (top panel of Fig.~\ref{fig:tspec})
makes it clear that the spectrum of energy is not contaminated by the
spectrum of the forcing function, as the latter only contains very low
frequencies, at or below $1/t^*$; this may not be the case with a
stochastic forcing function as the one used by \cite{nig04}.  This
underlines the role of turbulence in providing the high frequencies which
can resonate in the loop, viewed as a cavity for Alfv\'en waves, even in the
absence of an external driver at these frequencies.

\subsubsection{Auto-correlations}
\label{sec:autocorr}

The correlation time of the energy time series is a few dozens of the loop
crossing times by the Alfv\'en waves (Fig.~\ref{fig:ccorr}), consistent with
the slow evolution of the energy that we have already noted.  The
correlation time of the dissipation power time series is shorter, but still
longer than the wave crossing time, as an effect of the weakness of the
intermittent non-linear interactions between counter-propagation wavepackets
and of the global long-term fluctuations of the spectrum (including the
dissipation range) noted in Sect.~\ref{sec:spfluct}.

\subsubsection{Cross-correlations}
\label{sec:crosscorr}

\begin{figure}[tp]
  \plotone{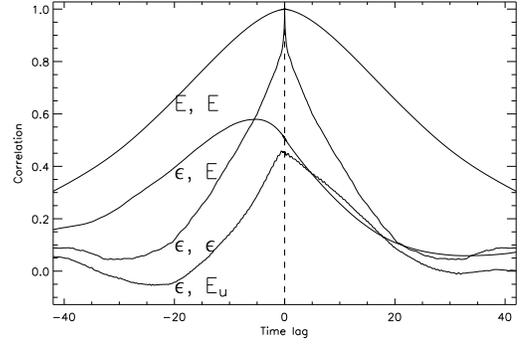}
  \caption{Auto-correlation functions and cross-correlation functions of
    time series of energy and energy dissipation power. $E$ is energy, $E_u$
    is kinetic energy and $\epsilon$ is the dissipation power.}
  \label{fig:ccorr}
\end{figure}

A common goal when studying solar flares and space weather is to find
precursors of flares, so as to get previsions of possible solar-terrestrial
events \citep[e.g.][]{abr02,abr05,geo05}. With this heating model we can use
the cross-correlations between time series to investigate which time series
react first, and what kind of observations would be helpful when predicting
flares.  We have extended the study of \citet{nig05}, who show that in some
events the kinetic energy begins to grow just before the start of a
dissipation event, by doing a systematic correlation study between all
energy and dissipation time series (kinetic, magnetic and total).  Figure
\ref{fig:ccorr} shows the cross-correlations between the dissipation
$\epsilon$ and the energy $E$ time series, as well as between the
dissipation $\epsilon$ and the kinetic energy $E_u$ time
series\footnote{Other correlations of pairs of time series of kinetic,
  magnetic or total energy or dissipations have not been plotted because
  they are very similar to either of the plotted correlations, as
  $\epsilon_u \ll \epsilon_b \approx \epsilon$ and $E_u \ll E_b \approx
  E$.}.  Both cross-correlation functions show delays of the dissipations
compared to the energy: the dissipation lags of approximately 5 time units
compared to the total energy, and of approximately 0.5 time units compared
to the kinetic energy (this last result is a confirmation of the result
obtained in \citealt{nig05}). Thus diagnostics methods based on the total
energy or including the magnetic fields measurements may provide more useful
results for space weather prediction than methods based on the velocity
field alone.  However, in both cases the delays involved are short, of the
order of a minute at best.

\subsection{Parametric study of dissipation power}
\label{sec:param}

\begin{figure}[tp]
  \plotone{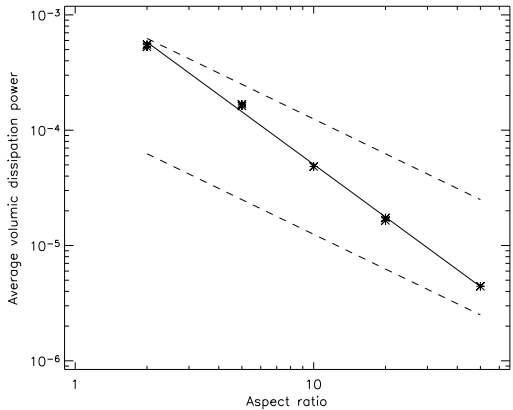}
  \caption{Average power of energy dissipation per unit volume (model
    dimensionless units) \vs aspect ratio $a$, for a fixed loop width
    $\ell=L/a=0.1$ and external field $b_o=1$. The power-law fit (plain
    line) has a slope $-1.52$. The two dashed lines represent a
    dissipatation power per unit surface of $10^2$ and $10^3\unit{W\cdot
      m^{-2}}$ respectively.}
  \label{fig:arde}
\end{figure}

Using the same runs as in Sect.\ref{sec:mken} (a set of loop with different
aspect ratios for a fixed width), we compute the average energy dissipation
in a stationary state, and we plot it versus the aspect ratio
(Fig.~\ref{fig:arde}): the scaling of the energy dissipation power per unit
volume is approximately in $a^{-3/2}$. This scaling can be compared to the
different models listed in \citet{man00} and corresponds to heating of MHD
in 2 dimensions.

As the slope of this power-law scaling is steeper than $-1$, shorter loops
are more efficient in dissipation power per unit surface. This can be
explained by the fact that Alfv\'en wavepackets that reflect on the loop
footpoints interact more frequently in a short loop than in a long loop; as
a matter of fact, simulations done when varying the Alfv\'en speed show that
the average dissipation power also increases when the Alfv\'en speed
increases.  Assuming that the physical units of the model (see
Sect.\ref{sec:scales} are $10\unit{Mm}$, $5\unit{s}$ and $10^9\unit{kg}$,
yielding $\ell=1\unit{Mm}$, $b_{\parallel}=2\unit{Mm\cdot s^{-1}}$ and
$\rho_0=10^{-12}\unit{kg\cdot m^{-3}}$, we get dissipation powers per unit
surface between $10^2\unit{W\cdot m^{-2}}$ for large aspect ratios and
$10^3\unit{W\cdot m^{-2}}$ for small aspect ratios. These values would be
sufficient to heat the quiet corona \citep{withb77}. Note however that they
also depend on the physical properties $b_{\parallel}$ and $\rho_0$ that we
have assumed for the loop.  Another series of runs has been performed to
explore the influence of $b_{\parallel}$ on the heating, and it gives
$\epsilon_{\mathrm{S}} \propto b_{\parallel}^{1.77}$; this translates the
fact that wavepackets interact more frequently when the Alfv\'en speed is
higher, leading to more dissipation.  These both fits, combined with a
dimensional analysis on the variables $\epsilon_{\mathrm{S}}$ (dissipation
power per unit surface), $\rho_0$ (mass density), $b_{\parallel}$ (Alfv\'en
speed), $u_f$ (forcing speed) and $a$ (aspect ratio), give:
\begin{equation}
  \label{eq:dissscal}
  \epsilon_{\mathrm{S}} = \frac{10^{2.22}}{a^{0.52}}
  \left(\frac{\rho_0}{10^{-12}}\right)
  \left(\frac{b_{\parallel}}{10^6}\right)^3
  \left(\frac{10^3 u_f}{b_{\parallel}}\right)^{1.23}
\end{equation}
for the dissipation power per unit surface in S.I. units
($\unit{W\cdot m^{-2}}$), as a function of the other variables in S.I. units
($\unit{kg\cdot m^{-3}}$ and $\unit{m\cdot s^{-1}}$).

\section{Conclusions}

We have presented the \shellatm\ model, which is a generalization of
shell-models \citep{giu98} with propagation of Alfv\'en waves along a \Bo\
field, with the further possibility of introducing a longitudinal
stratification of physical properties (\Bo, mass density, flux tube
expansion factor).  Although the model is simple and includes only
simplified physical processes, it has a very interesting complex non-linear
dynamics and it is fast enough to get statistics of its fields and of the
heating it produces: the simplifications we made allow to explore other
properties than those accessible to classical direct numerical simulations
(DNS).  While it is not meant to replace and cannot replace DNS, for example
because of the lack of three-dimensional information on field-line
topologies, it partially fills the huge gap between the Reynolds numbers in
DNS and in the real corona: although Reynolds numbers reached in the model,
of the order of $10^6$, are still lower than the ones expected in the
corona, this already represents an outstanding progress compared to DNS.
Furthermore this allows to explore regimes of MHD turbulence that are not
accessible to DNS, for instance this allows for intermittency to appear in
turbulence while having a complete description of the spectra and of the
non-linear interactions in the perpendicular direction.  On the other hand,
having even higher Reynolds numbers, and thus smallest scales, may require
to take into account non-MHD effects, like kinetic effects, that can still
be neglected in this model.

The model has been used in this paper in the case of a magnetic loop in the
solar corona, in which the physical properties of the medium (namely the
external longitudinal magnetic field \Bo\ and the mass density) are assumed
to be uniform along the loop.  In this case, and thanks to the
above-mentioned characteristics of the model, we could show that this model
loop displays: a dynamics over a very wide range of spatial and temporal
scales (4 to 5 orders of magnitude); spectra which are formed by a local
cross-scale energy flux, and which have a wide inertial range in either
direction, perpendicular or parallel to the external magnetic field; a clear
anisotropy between the parallel and perpendicular spectra, which could be
compatible with the ``critical balance'' phenomenology; a scaling of the
average ratio of the magnetic energy over the kinetic energy consistent with
RMHD; a heating function with multiple spatial and temporal scales; a flat
longitudinal profile of the average dissipation power (although this may be
dismissed by further simulations, with non-uniform physical properties of
the medium along the loop); a spiky, and statistically intermittent time
series of energy dissipation power; power-law distributions of the
characteristics (peak energy, total energy, duration, waiting-times) of
``events'' extracted from the time series of dissipation power;
finite-lifetime packets of resonant frequencies in the time series of
energy, but whose frequencies are shifted from the harmonics of the linear
resonant frequencies because of the non-linearities; long-range time
correlations in time series; a delay of the dissipation time series compared
to some energy time series; an average dissipation power that scales with
the loop parameters, and that could be sufficient to sustain the high
coronal temperatures.  As discussed in the text, some of these results
confirm or complete the results of a similar model \citep{nig04,nig05}.

Further directions for the study of the solar corona using this model
include taking advantage of the possibility of modelling non-uniform regions
to (1) allow for density gradients in a coronal loop, to seek for the
preferred locations of coronal heating, and (2) study a magnetically open
region such as a coronal hole.  In order to get diagnostics that can be
compared to observations, this heating model can be coupled to the
thermodynamics of a loop (including the cooling by conduction and radiation)
upon which forward-modelling of coronal spectral lines may be carried out.
We also believe that this model can be used in other heliospheric and
astrophysical applications where MHD applies and where there is a strong
dominant magnetic field (see Sect.~\ref{sec:avail} in Appendix for code
availability).

\acknowledgements
  The authors acknowledge financial support from European Union grant
  HPRN-CT-2001-00310 (TOSTISP network).  They thank the referee for
    the suggested improvements to the manuscript.  Discussions with
  S.~Galtier and collaboration with A.~Verdini when developing the numerical
  code are greatly acknowledged.  Part of this work was done while attending
  the spring 2005 programme Grand Challenge Problems in Computational
  Astrophysics, at Institute for Pure and Applied Mathematics, UCLA, Los
  Angeles, CA. Wavelet software was provided by C.\ Torrence and G.\ Compo,
  and is available at \url{http://paos.colorado.edu/research/wavelets/}.
  Computations were done on Linux clusters at Arcetri Observatory, CINECA
  and JPL.

\appendix
\section{The numerical code}
\label{sec:num}

\subsection{Numerical schemes}
\label{sec:numsch}

The time advancement of the non-linear terms of the shell-models is
performed by a third-order Runge-Kutta scheme. The Alfv\'en-wave propagation
is done by the Fromm numerical scheme \citep{fromm68}.  Finally, the
dissipation terms of the shell-models can be computed by an implicit scheme.
This allows to relax the CFL condition on \tnu\ and thus to fully resolve
the dissipation range of the spectrum at no further computational cost.

\subsection{Parallelization and parallel efficiency}
\label{sec:parall}

The \shellatm\ model is parallelized using MPI, by simply distributing the
planes over the processors. Communications are mainly needed for the
propagation of the Alfv\'en waves between the domains corresponding to the
different processors, and for the output. The resulting parallelization
efficiency is good, and is even close to the ideal parallelization
efficiency (up to hundreds of processors for $n_z=10^4$) thanks in
particular to effects due to the cache size (when the number of processors
grows, the local data get small enough to fit entirely in the level-2 memory
cache of each processor).

\subsection{Architecture of the code and availability}
\label{sec:avail}

The \shellatm\ code is modular and can be adapted to a large variety of
physical systems. Different models for the non-linearities and different
numerical schemes can be chosen. We believe that the code can be useful for
the community and we have thus released it under the GNU GPL licence. The
code and its manual can be found at\\
\url{http://www.arcetri.astro.it/\~eric/shell-atm/codedoc/}.

\end{document}